\documentclass[
  aps,
  prb,
  twocolumn,
  preprintnumbers,
  floatfix,
  showpacs,
  amsmath,
  amssymb,
  superscriptaddress,
]{revtex4-1}

\usepackage{amsmath,amssymb}
\usepackage{graphicx}
\usepackage{color}
\usepackage{dcolumn}
\usepackage{subfigure}
\usepackage{times}

\newcommand{\VBM}{\text{VBM}}
\newcommand{\Pb}{\text{Pb}}
\newcommand{\Ti}{\text{Ti}}
\newcommand{\Fe}{\text{Fe}}
\newcommand{\Cu}{\text{Cu}}

\renewcommand{\O}{\text{O}}
\newcommand{\PT}{\text{PbTiO$_3$}}

\newcommand{\grad}{\ensuremath{^\circ}}
\renewcommand{\deg}{\grad}

\newcommand{\dnspin}[0]{\ensuremath\Delta n_{\upharpoonleft\downharpoonright}}
\newcommand{\mydot}[0]{\scriptscriptstyle{\bullet}}

\newcommand{\eV}{\text{eV}}

\newcommand{\tbhd}[1]{\textbf{#1}}
\newcommand{\fig}[1]{Fig.~\ref{#1}}
\newcommand{\tab}[1]{Table~\ref{#1}}
\newcommand{\sect}[1]{Sec.~\ref{#1}}
\newcommand{\eq}[1]{(\ref{#1})}

\begin{document}

\preprint{
  published as Phys. Rev. B {\bf 76}, 174116 (2007) \hspace{156pt}
  Copyright (2007) The American Physical Society
}

\title{
  Association of oxygen vacancies
  with impurity metal ions in lead titanate
}

\author{Paul Erhart}
\affiliation{
  Institut f\"ur Materialwissenschaft,
  Technische Universit\"at Darmstadt,
  64287 Darmstadt, Germany
}
\affiliation{
  Lawrence Livermore National Laboratory,
  Chemistry, Materials and Life Sciences Directorate,
  Livermore, California, 94551
}
\author{R\"udiger-Albert Eichel}
\affiliation{
  Eduard-Zintl-Institut,
  Technische Universit\"at Darmstadt,
  64287 Darmstadt, Germany
}
\author{Petra Tr\"askelin}
\affiliation{
  Accelerator Laboratory,
  University of Helsinki,
  00014 Helsinki, Finland
}
\affiliation{
  Department of Chemical Engineering and Materials Science,
  University of California at Davis, California, 95616
}
\author{Karsten Albe}
\affiliation{
  Institut f\"ur Materialwissenschaft,
  Technische Universit\"at Darmstadt,
  64287 Darmstadt, Germany
}

\begin{abstract}
Thermodynamic, structural and electronic properties of isolated
copper and iron atoms as well as their complexes with oxygen vacancies in
tetragonal lead titanate are investigated by means of first
principles calculations. Both dopants exhibit a strong chemical driving
force for the formation of $M_{\Ti}-V_{\O}$ ($M$ = Cu, Fe) defect
associates. The most stable configurations corresponds to a local
dipole aligned along the tetragonal axis parallel to the spontaneous
polarization. Local spin moments are obtained and the calculated spin
densities are discussed. The calculations provide a simple and
consistent explanation for the experimental findings. The results are
discussed in the context of models for degradation of ferroelectric
materials.
\end{abstract}

\pacs{
  61.72.Ji, 
  71.15.Mb, 
  77.22.Ej, 
  77.84.Dy  
}

\maketitle

\section{Introduction}

Piezoelectric materials transform electric signals into mechanical
response and vice versa. They are widely used in sensors and actuators
as well as in non-volatile memory devices. The properties of
piezoelectrics and their behavior in the presence of electric fields
are intimately linked to the defect structures (point defects, domain
walls, grain boundaries) present in these materials.

Defect dipoles, for instance, formed by the association of oxygen
vacancies with metal impurities affect the switching behavior as well
as aging and fatigue of piezoelectric materials. \cite{WarTutDim95}
Their existence has been shown both experimentally \cite{WarDimPik95,
  WarTutDim95, MesEicDin04, MesEicKlo05} and theoretically.
\cite{PoyCha99a} However, a comprehensive microscopic picture of the
electronic, structural and kinetic properties of these defect
associates is still lacking. Such detailed knowledge is, however, a
prerequisite for understanding  electrical ``aging'' and ferroelectric
``fatigue'' as exemplified by the model by Arlt and
Neumann,\cite{NeuArl87, ArlNeu88} which relates these macroscopic
degradation phenomena to microscopic processes. 
Since the time the first phenomenological models were developed
\cite{NeuArl87, ArlNeu88} quantum mechanical methods have become
available which are both efficient and sufficiently reliable for
assessing the energetics of point defect and associates, directly. In
the past, several studies addressed the properties of intrinsic point
defects with particular attention to oxygen vacancies. \cite{ParCha98,
  PoyCha00a, CocBur04, Par03, HeVan03, ZhaWuLu06} The role of dopants
and impurities and the formation of defect dipoles has also been
elucidated \cite{PoyCha99a, PoyCha00b, MesEicDin04, MesEicKlo05} but
important aspects such as the dependence of the formation energies on
the Fermi level and the charge state have not been investigated in
sufficient detail. Also, a comparative study of the different possible
metal impurity-oxygen vacancy configurations, which could serve as
input for the defect models referenced above, is not available
yet. Such information is, however, crucial both for understanding the
generic properties of different impurities/dopants and for identifying
properties which differentiate them, in particular with regard to
their effect on aging and fatigue. These considerations motivate the
present work, in which density functional theory calculations are
employed to study the energy surface for unbound oxygen vacancies and
oxygen vacancies complexed with Fe or Cu impurities in tetragonal lead
titanate. The results are discussed in the context of existing
defect models and have been used to interpret recent electron spin
resonance experiments.\cite{EicErhTra07}

\section{Methodology}

\subsection{Computational details}

Quantum mechanical calculations based on density functional theory
(DFT) were carried out using the Vienna ab-initio simulation package
(\textsc{vasp}) which implements the pseudopotential-plane wave
approach. \cite{KreHaf93, KreHaf94, KreFur96a, KreFur96b} The
projector-augmented wave method was employed for representing
the ionic cores and core electrons. \cite{Blo94, KreJou99} In accord
with earlier calculations
(see e.g., Refs.~\onlinecite{ParCha98} and \onlinecite{MesEicKlo05})
the $5d$ electrons of Pb and the $3s$ and $3p$ electrons of Ti were
treated as semi-core states, while for Cu and Fe the valence
configuration included the $4d$, $3d$ and $3p$ electrons. All
calculations were performed using Gaussian smearing with a width of
$\sigma=0.2\,\eV$ and a $4\times 4\times 4$ Monkhorst-Pack
mesh for Brillouin zone sampling. \cite{MonPac76} Calculations were
carried out using the local spin density approximation (LSDA) to
account for the magnetic behavior of iron.\cite{Note1}
Some calculations were also repeated using the non-spin polarized
local density approximation (LDA). The difference between the LDA and
LSDA results allows to estimate the spin contribution for a particular
defect or charge state.

With this computational setup, we obtain a lattice
constant of $a_0=3.866\,\text{\AA}$ and an axial ratio of $c/a=1.05$
for the tetragonal phase of \PT\ in reasonable agreement with
experiment. The calculated band gap of 
1.47\,eV is considerably smaller than the experimental value but
consistent with the well known LSDA band gap error.
Since there is no unique procedure for correcting the band gap error (see
e.g, Refs.~\onlinecite{PerZhaLan05, JanVan06, ErhAlb07a}),  in the
present work, we abstain from any explicit corrections and discuss the
electronic transition levels relative to the calculated band edges
instead (see also our discussion in \sect{sect:VO}).

\begin{figure}
  \centering
  \includegraphics[width=0.86\linewidth]{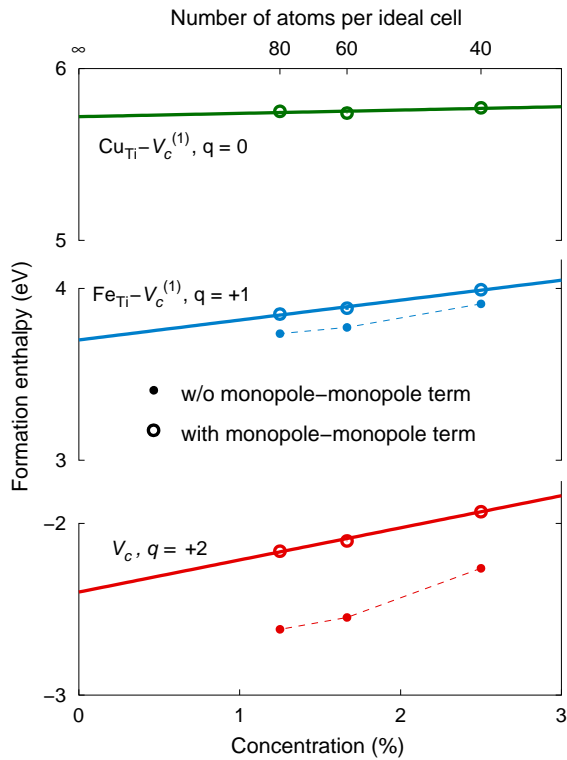}
  \caption{
    (Color online)
    Finite-size scaling of formation enthalpies to obtain formation
    energies in the limit of infinite dilution for three of the most
    important defects. The small filled and large open circles show
    the data before and after application of the monopole-monopole
    correction term.
 }
  \label{fig:finsize}
\end{figure}

In order to model the defect configurations, super cells containing
$2\times 2\times 2$ to $2\times 2\times 4$ unit cells equivalent to 40
to 80 atoms were employed. Each configuration was relaxed until the
maximum force was less than 15\,meV/\AA. For charged defects a
homogeneous background charge was added. Finite-size scaling was
employed in order to obtain the formation and binding energies at
infinite dilution. \cite{LenMozNie02, ErhAlbKle06} For the charged
defect calculations monopole-monopole interactions were explicitly
corrected \cite{MakPay95, LenMozNie02} using values of
$\epsilon_{11}^S=100$ and $\epsilon_{33}^S=34$ for the static
dielectric tensor, \cite{Note2}
while higher order terms were implicitly included by the finite-size
scaling procedure. \cite{LenMozNie02, ErhAlbKle06}
Both the effect of the monopole-monopole term and the finite-size
procedure are illustrated in \fig{fig:finsize} for three of the most
important defects which demonstrates that the finite-size scaling
approach works reliably in the present case. The extrapolation errors
of the finite-size scaling procedure are given in
Tables~\ref{tab:eform_VO}, \ref{tab:eform_CuTi}, and
\ref{tab:eform_FeTi}. These values are a measure for the quality of
the finite-size scaling procedure only, and do not include other
errors intrinsic to DFT.

\subsection{Formation energies}

The formation energy of an intrinsic defect in charge state, $q$,
depends on the relative chemical potentials of the constituents,
$\Delta\mu_i$, and the Fermi level (electron chemical potential),
$\mu_e$, according to \cite{QiaMarCha88, ZhaNor91}
\begin{align}
  \Delta E_D^f
  &= (E_D - E_H) + q (E_{\VBM} + \mu_e) \nonumber\\
  &\quad - \sum_i \Delta n_i (\mu_i^{bulk} + \Delta\mu_i)
  \label{eq:eform}
\end{align}
where $E_D$ is the total energy of the defective system, $E_H$ is
the total energy of the perfect reference cell, $E_{\VBM}$
is the position of the valence band maximum, $\Delta n_i$ denotes
the difference between the number of atoms of type $i$ in the
reference cell with respect to the defective cell, and $\mu_i^{bulk}$
is the chemical potential of the reference phase of atom type
$i$.
In the present context we are not concerned with the effect of the
chemical environment ($\Delta\mu_i$ in equation \eq{eq:eform}) but
focus on the transition levels and the binding energies between oxygen
vacancies and metal impurities/dopants (see equation \eq{eq:ebind}
below), both of which are independent of the chemical potentials.

The binding energy for the metal impurity-oxygen vacancy
($M_{\Ti}-V_{\O}$) associates considered below is calculated as the
difference between the formation energies of the defect complex and
the isolated defects
\begin{align}
  E_b
  &= \Delta E_D^f[M_{\Ti}-V_{\O}]
  -  \Delta E_D^f[M_{\Ti}]
  -  \Delta E_D^f[V_{\O}].
  \label{eq:ebind}
\end{align}
The binding energy depends on the Fermi level due to the Fermi level
dependence of the formation energies. For a given value of $\mu_e$ the
difference in equation \eq{eq:ebind} is calculated between the most
stable charge states for each defect involved.
By insertion of equation \eq{eq:eform} into \eq{eq:ebind} one can show
that the terms with $\Delta n_i$ cancel, whence $E_b$ is independent
of the chemical potentials, $\Delta\mu_i$. Following the sign
convention introduced via equation \eq{eq:ebind}, negative values for
$E_b$ imply that the system gains energy by association, whereas
positive values imply an effective repulsion between the isolated
defects.

\section{Results}

\subsection{Isolated oxygen vacancies}
\label{sect:VO}

\begin{table}[b]

  \caption{
    Formation energies (eV) for unbound oxygen vacancies in tetragonal
    lead titanate for a Fermi level at the valence band maximum
    ($\mu_e=0\,\eV$) and metal-rich conditions
    ($\Delta\mu_{\Pb}=\Delta\mu_{\Ti}=0\,\eV$). The finite-size
    scaling extrapolation errors are given in brackets in the last
    column.
  }
  \label{tab:eform_VO}

  \centering

  \newcolumntype{Q}[0]{>{(}d<{)}}
  \newcommand{\spread}[1]{\multicolumn{1}{c}{#1}}
  \newcommand{\spreadd}[1]{\multicolumn{2}{c}{#1}}

  \begin{tabular}{ldddddQ}
    \hline\hline

    \tbhd{Defect}
    & 
    & \spread{$\boldsymbol{40}$}
    & \spread{$\boldsymbol{60}$}
    & \spread{$\boldsymbol{80}$}
    & \spreadd{\tbhd{extrapolated}} \\
    
    \hline
    
    $V_{ab}$
    & 0  &   0.96 &   0.95 &   0.95 &   0.95 &  <0.01 \\
    & +1 &  -0.40 &  -0.43 &  -0.43 &  -0.47 &   0.01 \\
    & +2 &  -1.64 &  -1.62 &  -1.62 &  -1.60 &  <0.01 \\[6pt]

    $V_c$
    & 0  &   1.07 &   0.87 &   0.70 &   0.35 &   0.08 \\
    & +1 &  -0.59 &  -0.88 &  -0.96 &  -1.35 &   0.06 \\
    & +2 &  -1.93 &  -2.10 &  -2.16 &  -2.40 &   0.03 \\

    \hline\hline
  \end{tabular}

\end{table}

The crystallographic structure of a tetragonal perovskite possesses
two symmetrically inequivalent oxygen sites which are located within
the $ab$-plane and along the $c$-axis, respectively. Based on DFT
calculations, Park and Chadi \cite{ParCha98} identified one possible
vacancy configuration along the $c$-axis ($V_c$) and two distinct
vacancies configurations within the $ab$-plane ($V_{ab}^{sw}$,
$V_{ab}^{ud}$). The latter two configurations differ in the
orientation of the polarization above and below the vacancy plane. In
the present work we have found the ``up-down'' configuration
($V_{ab}^{ud}$, see Ref.~\onlinecite{ParCha98} for nomenclature), in
which the polarization vectors are oriented head-to-head in the
vacancy plane, to be highly unstable with respect 
to the ``switchable'' configuration ($V_{ab}^{sw}$), which is
characterized by a uniform orientation of the polarization
vector. Although Park and Chadi did not provide a quantitative energy
comparison for $V_{ab}^{sw}$ and $V_{ab}^{ud}$ due to missing data for
the 180\deg-domain wall energy, they estimated the $V_{ab}^{sw}$
configuration to be lower in energy. \cite{ParCha98} For these
reasons, we focus on the $V_{ab}^{sw}$ configuration which for the
sake of brevity in the following is denoted $V_{ab}$.

Table \ref{tab:eform_VO} summarizes the formation energies for unbound
oxygen vacancies obtained for extreme metal-rich conditions
($\Delta\mu_{\Pb} = \Delta\mu_{\Ti} = 0\,\eV$).
In all charge states the $c$-type vacancy is energetically preferred
over the $ab$-type vacancy in agreement with the calculations in
Ref.~\onlinecite{ParCha98}. For infinite dilution (extrapolated values
in \tab{tab:eform_VO} the energy difference is as large as 0.8\,eV for
$q=+2$ which is on the order of magnitude of the energy barrier for
migration.\cite{ErhTraAlb07}

\begin{figure}
  \centering
  \includegraphics[scale=0.7]{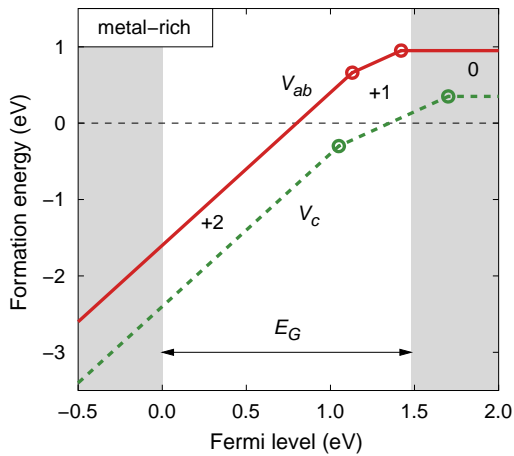}
  \caption{
    (Color online)
    Variation of the formation energy of unbound oxygen vacancies
    with the Fermi level for extreme metal-rich conditions
    ($\Delta\mu_{\Pb} =\Delta\mu_{\Ti} =0\,\eV$). The white area
    corresponds to the calculated band gap. The numbers indicate the
    defect charge state.
  }
  \label{fig:eform_VO}
\end{figure}

The variation of the formation energy with the Fermi level is shown in
\fig{fig:eform_VO}. Practically over the entire band gap oxygen
vacancies are doubly positively charged. For the given chemical
conditions the formation energies are negative which indicates that
the material is unstable for these chemical potentials
($\Delta\mu_{\Pb} =\Delta\mu_{\Ti} =0\,\eV$).
If the oxygen partial pressure is raised (i.e., $\Delta\mu_{\Ti}$
and $\Delta\mu_{\Pb}$ become less negative while $\Delta\mu_{\O}$
becomes more negative), the oxygen vacancy formation energy
increases and the stability condition is fulfilled.

On first sight the present finding is at variance with the
calculations in Ref.~\onlinecite{Par03} in which the +2/0 transition
level of the oxygen vacancy is located in the middle of the band
gap. This finding is, however, a consequence of the corrections
applied in that work: In order to correct for the
the band gap error the conduction band minimum was rigidly
shifted upwards \emph{without} correcting the
formation energies. If this shift is omitted the results of
Ref.~\onlinecite{Par03} and the findings of the present work are
consistent with each other. The approach in Ref.~\onlinecite{Par03}
ignores the fact that in the case of singly charged
($V_{\O}^{\mydot}$) and neutral oxygen vacancies
($V_{\O}^{\times}$) the excess electron(s) occupy conduction band
states. In order to maintain internal consistency, a correction term
\cite{PerZhaLan05} on the order of $z_e \Delta E_G$ would have to be
added to the formation energies of $V_{\O}^{\mydot}$ and
$V_{\O}^{\times}$, where $z_e$ is the number of occupied
conduction band states and $\Delta E_G$ denotes the difference between
the calculated and the experimental band gap. \cite{Note3}
If these energy terms are included, the +2/+1 (+2/0) transition is no
longer located in the middle of the (experimental) band gap but near
the conduction band minimum. This is equivalent to the location of the
transition level with respect to the calculated CBM if no correction
is applied.

\subsubsection{Isolated copper impurities}

\begin{table}

  \caption{
    Formation energies (LSDA) in units of eV for isolated and
    complexed Cu-impurities on Ti-sites ($\Cu_{\Ti}$) in tetragonal
    lead titanate for a Fermi level at the valence band maximum
    ($\mu_e=0\,\eV$) and metal-rich conditions ($\Delta\mu_{\Pb}
    =\Delta\mu_{\Ti} =\Delta\mu_{\Cu} =0\,\eV$). The finite-size
    scaling extrapolation errors are given in brackets in the last but
    one column.
    $\dnspin$ denotes the difference between the number of electrons
    in spin-up and spin-down states.
  }

  \label{tab:eform_CuTi}

  \centering

  \newcolumntype{Q}[0]{>{(}d<{)}}
  \newcommand{\spread}[1]{\multicolumn{1}{c}{#1}}
  \newcommand{\spreadd}[1]{\multicolumn{2}{c}{#1}}

  \renewcommand{\baselinestretch}{1.0}
  \small

  \begin{tabular}{lddddddQd}
    \hline\hline

    \multicolumn{2}{l}{\tbhd{Defect}}
    & \spread{$\boldsymbol{\dnspin}$}
    & \spread{$\boldsymbol{40}$}
    & \spread{$\boldsymbol{60}$}
    & \spread{$\boldsymbol{80}$}
    & \spreadd{\tbhd{extrapolated}}
    \\
    
    \hline
    
    \multicolumn{4}{l}{
      $\Cu_{\Ti}$
    } \\
    & -3 & - &  11.81 &  12.04 &  12.06 &  12.35 &   0.10 \\
    & -2 & 1 &  10.16 &  10.28 &  10.30 &  10.46 &   0.05 \\
    & -1 & 2 &   9.74 &   9.79 &   9.81 &   9.88 &   0.01 \\
    & 0  & - &   9.97 &   9.95 &  10.07 &  10.11 &   0.14 \\[6pt]

    \multicolumn{4}{l}{
      $\Cu_{\Ti}-V_{ab}$
    } \\
    & -2 & - &   9.22 &   9.32 &   9.32 &   9.45 &   0.06 \\
    & -1 & 0 &   7.23 &   7.36 &   7.41 &   7.60 &   0.01 \\
    & 0  & 1 &   6.43 &   6.45 &   6.48 &   6.52 &   0.02 \\
    & +1 & 0 &   6.67 &   6.62 &   6.60 &   6.52 &   0.01 \\[6pt]

    \multicolumn{4}{l}{
      $\Cu_{\Ti}-V_c^{(1)}$
    } \\      
    & -2 & - &   8.83 &   8.91 &   8.91 &   9.01 &   0.04 \\
    & -1 & 0 &   6.86 &   6.94 &   6.99 &   7.11 &   0.01 \\
    & 0  & 1 &   5.77 &   5.74 &   5.75 &   5.72 &   0.03 \\
    & +1 & 0 &   6.02 &   5.94 &   5.91 &   5.80 &   0.01 \\[6pt]

    \multicolumn{4}{l}{
      $\Cu_{\Ti}-V_c^{(2)}$
    } \\      
    & -2 & - &   9.33 &   9.57 &   9.69 &  10.04 &  <0.01 \\
    & -1 & 0 &   7.29 &   7.59 &   7.77 &   8.24 &   0.03 \\
    & 0  & 1 &   6.22 &   6.43 &   6.59 &   6.93 &   0.06 \\
    & +1 & - &   6.44 &   6.64 &   6.80 &   7.14 &   0.07 \\[6pt]

    \multicolumn{4}{l}{
      $\Cu_{\Ti}-V_c^{(3)}$
    } \\
    & -2 & - &        &  10.00 &  10.01 &  10.04 & \spread{} \\
    & -1 & 0 &        &   7.98 &   8.04 &   8.22 & \spread{} \\
    & 0  & 1 &        &   6.76 &   6.75 &   6.74 & \spread{} \\
    & +1 & - &        &   6.97 &   6.89 &   6.68 & \spread{} \\[6pt]

    \multicolumn{4}{l}{
      $\Cu_{\Ti}-V_c^{(4)}$
    } \\
    & -2 & - &        &        &  10.21 &        & \spread{} \\
    & -1 & 0 &        &        &   8.26 &        & \spread{} \\
    & 0  & 1 &        &        &   6.98 &        & \spread{} \\
    & +1 & - &        &        &   7.14 &        & \spread{} \\
    
    \hline\hline
    
  \end{tabular}

\end{table}

\begin{figure}
  \centering
\includegraphics[scale=0.7]{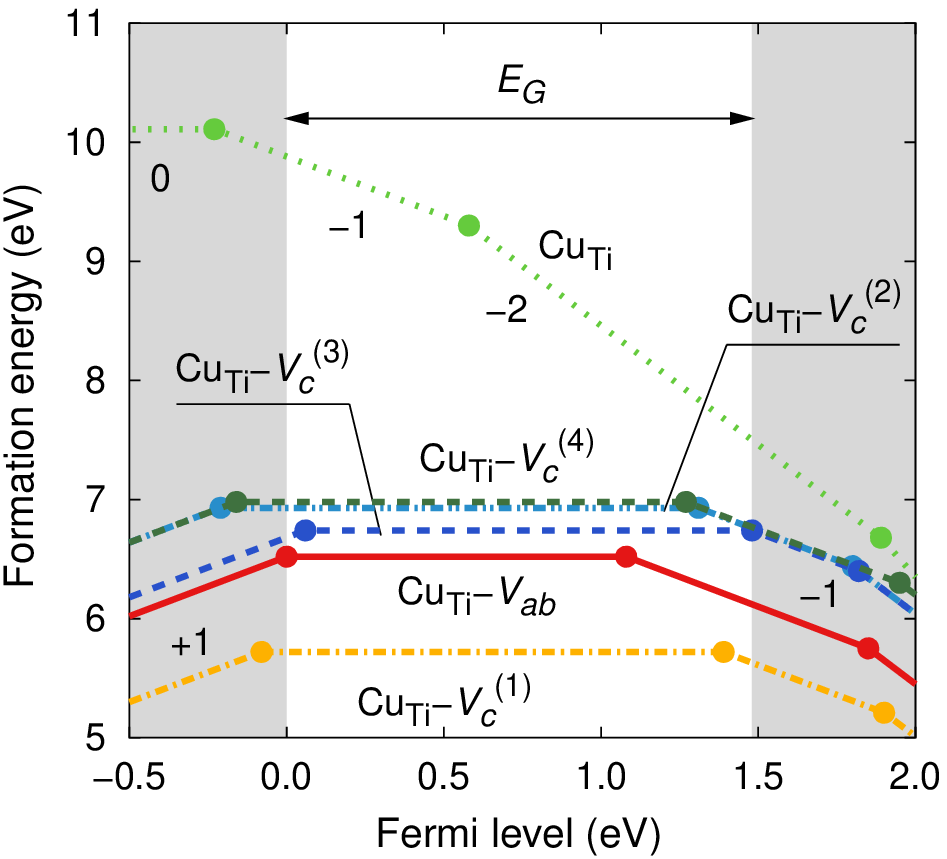}
  \caption{
    (Color online)
    Fermi level dependence of formation energies of free and complexed
    Cu impurities. Parallel lines imply equal charge states as
    indicated in the plot. The white stripe indicates the (calculated)
    band gap while the gray stripes indicate the valence and
    conduction bands, respectively. Extrapolated formation energies
    from \tab{tab:eform_CuTi} were used except for $\Cu_{\Ti}-V_c^{(4)}$
    for which the data obtained with 80-atom cells are shown.
  }
  \label{fig:eform_CuTi}
\end{figure}

Since there is experimental evidence that copper impurities in lead
titanate preferentially occupy Ti-sites \cite{EicKunHof04,
  EicDinKun05, EicMesDin05} only the substitutional position was
considered ($\Cu_{\Ti}$) in the present calculations. The calculated
formation energies for different charge states of $\Cu_{\Ti}$ are
compiled in \tab{tab:eform_CuTi} for extreme metal-rich conditions
($\Delta\mu_{\Pb}= \Delta\mu_{\Ti}= \Delta\mu_{\Cu}=
\Delta\mu_{\Fe}=0\,\eV$). They can be used to derive the Fermi level
dependence of the defect formation energy as demonstrated in
\fig{fig:eform_CuTi} which shows that uncomplexed Cu-impurities can
occur in charge states $q=-1$ and $-2$. The transition between these
two charge states is located $\sim\!0.2\,\eV$ below the middle of the
calculated band gap.

\begin{figure}
  \centering
  \includegraphics[width=0.9\linewidth]{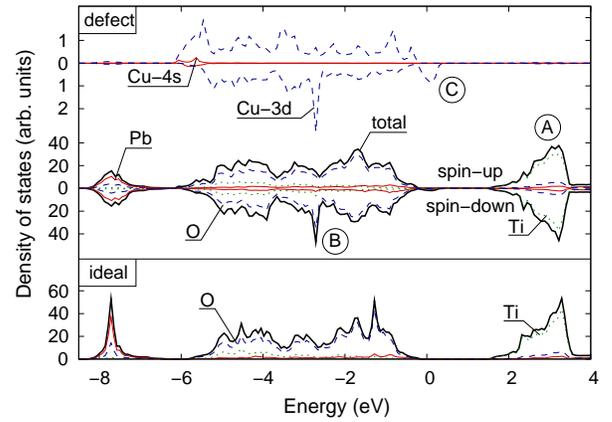}
  \caption{
    (Color online)
    Spin-decomposed total and partial density of states for neutral
    uncomplexed Cu impurities in lead titanate. The partial densities
    of states for the Cu atom have been scaled by a factor of twenty
    and are shown separately in the upper part of the figure.
    The density of states for the ideal crystal is shown
    below. In the latter case the density of states is independent of
    the spin orientation
    A Gaussian filter with $\sigma=0.05\,\eV$ was applied to optimize
    the visualization.
  }
  \label{fig:spin_CuTi}
\end{figure}

In the defect chemistry of oxides, often a fully ionic picture is
adopted e.g., in Cu-doped lead titanate $\Cu^{2+}$ ions are usually
assumed to replace $\Ti^{4+}$ ions with two defect electrons being
associated with the defect site. In Kr\"oger-Vink notation the
resulting defect is written as $\Cu_{\Ti}^{''}$. Accordingly,
copper-impurities should act as acceptors and create holes in the
valence band.

The present quantum-mechanical calculations are free of {\it a-priori}
assumptions with respect to the oxidation state of a defect:
Different charge states, $q$, are obtained by varying the number of
electrons in the {\it entire} system. In order to provide a connection
between the ionic picture outlined above and the charge states given
in \tab{tab:eform_CuTi} and \fig{fig:eform_CuTi}, one must therefore
analyze the electronic structure of the defect.

\begin{figure}
  \centering
\includegraphics[width=0.96\linewidth]{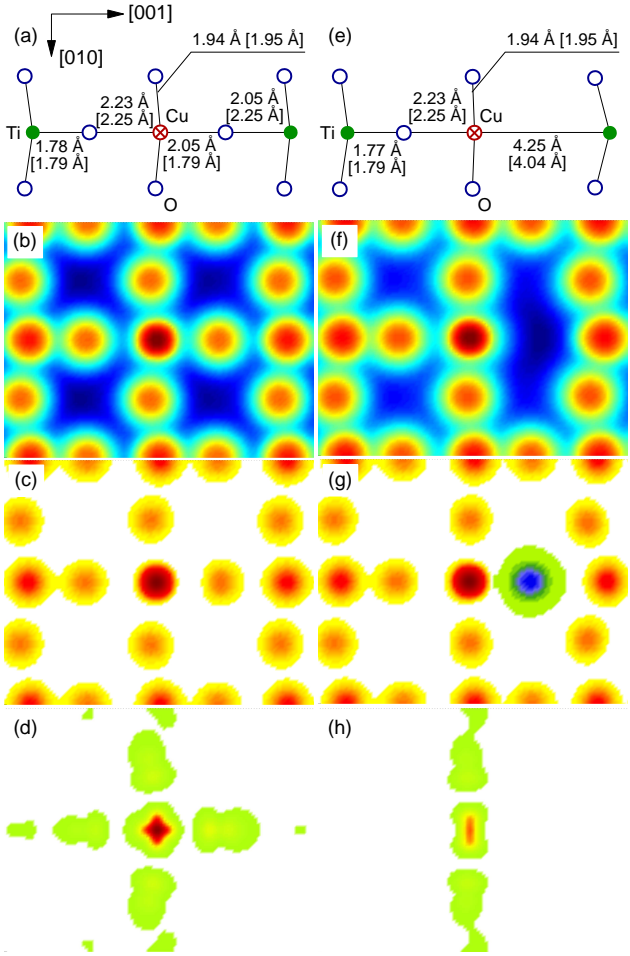}
  \caption{
    (Color online)
    Isolated copper impurity in charge state $q=-1$ (a-d) and copper
    impurity-oxygen vacancy complex in charge state $q=0$ (e-h).
    (a,e) Position of atoms in the (100)-plane containing the defect,
    numbers given in square brackets give the equivalent distances in
    the defect-free crystal;
    (b,f) total charge density, a logarithmic scale has been chosen for
    the charge density in order to enhance the features,
    (c,g) difference between total charge density of defective cell
    and defect-free cell, and
    (d,h) spin densities.
  }
  \label{fig:dens_CuTi}
  \label{fig:dens_CuTi_VO}
\end{figure}

The partial and total densities of states for a neutral ($q=0$) copper
impurity in comparison with the ideal (defect-free) system are shown
in \fig{fig:spin_CuTi}. The bottom of the conduction band is
predominantly composed of Ti--3d orbitals (feature A in
\fig{fig:spin_CuTi}). Since in this energy range the contribution
Cu--3d orbitals is practically zero, copper substitution leads to an
effective reduction of the density of states at the bottom of the
conduction band. With regard to the top
of the valence band, the most obvious feature is a hybridization of
Cu--3d orbitals with O--2p orbitals (feature B in \fig{fig:spin_CuTi})
while the contribution of Cu-4s and Cu--3p states is negligible. In
addition, a Cu-induced state emerges attached to the valence band edge
(feature C in \fig{fig:spin_CuTi}).
As electrons are added to the system, this defect 
induced state is, however, shifted into the band gap because the
excess charge is localized in a small volume around the defect. As a
result the local potential at the copper atom is modified and an
energy offset of the electronic states of this defect is induced.

Integrating the density of states and the occupied levels up to the
valence band edge and including the defect induced peak, reveals three
empty levels (holes) for the neutral charge state. As electrons are
added to the system ($q<0$) these empty states are gradually
filled. For the charge state $-1$ there are two holes localized at the
Cu atom. The situation is thus equivalent to placing a $\Cu^{++}$ on a
Ti-site giving $\Cu_{\Ti}^{''}$, while for $q=-2$ there is one
unoccupied valence band level corresponding to $\Cu_{\Ti}^{'''}$
($\Cu^+$ on $\Ti^{4+}$). If the Fermi level is located in the middle
of the band gap, the most stable defect is $\Cu_{\Ti}^{'''}$.

The $-1/-2$ transition can be understood as a local reduction of
copper: The transition from charge state $q=-1$ to $q=-2$
is equivalent to the reaction $\Cu^{++} + e^- \rightarrow 
\Cu^+$. Experimentally, the redox potential for this reaction has been
measured in aqueous solution to be $+0.16\,\eV$ with respect to the
standard hydrogen electrode, which on an absolute energy scale is
located at 4.5\,eV. \cite{Tra90} Thus, the 
$\Cu^{++}/\Cu^+$ redox potential is 4.7\,eV. On the other hand, the
electron affinity of \PT\ is 3.5\,eV and the band gap is about 3.4\,eV
(Ref.~\onlinecite{RobChe99} and references therein) placing the
valence band maximum (VBM) 6.9\,eV below the vacuum level. In analogy,
the redox reaction occurs roughly in the middle of the band gap
(1.2\,eV below the CBM or 2.2\,eV above the VBM), which is in good
agreement with the present calculations which locate the $-1/-2$ (or
$\Cu^{++}/\Cu^{+}$) transition level near the mid gap. This analogy
has, however, merely a qualitative character.

In order to obtain a measure for the spin-polarization of copper
defects we calculated the difference between the number of electrons
in spin-up and spin-down states ($\dnspin$ in
\tab{tab:eform_CuTi}). In both relevant charge states ($q=-1$ and
$-2$) unpaired electrons are present in the system which allows to
probe the defect center spectroscopically.

The geometry and charge density of an isolated copper impurity in
charge state $q=-1$ is shown in \fig{fig:dens_CuTi}. In comparison
with the ideal crystal, copper substitution causes the $B$-site to
move towards the oxygen (100)-plane. For the charge state $q=-1$
($q=-2$) the relative displacement of the copper atom with respect to
its in-plane oxygen neighbors is 0.15\,\AA\ (0.19\,\AA) along $[001]$,
compared to 0.28\,\AA\ in the defect free crystal. Copper substitution
thus causes a relaxation of the $B$-site atom in the direction of the
nearest oxygen plane. The charge density plots show that the defect
charge (\fig{fig:dens_CuTi}(c)) as well as the spin-asymmetry
(\fig{fig:dens_CuTi}(d), also compare \fig{fig:spin_CuTi}) is
localized at the copper atom. In addition, the difference between the
spin-densities in the $(100)$-plane reveals a slight asymmetry with
respect to the tetragonal axis. Analysis of the electron and spin
density at farther atoms -- in particular the nearest Pb atoms in the
$(110)$-plane -- indicates that both of them are virtually unaffected
by the presence of the Cu impurity.

\subsubsection{Copper-vacancy complexes}
\label{sect:CuVO}

\begin{figure}
  \centering
\includegraphics[width=\linewidth]{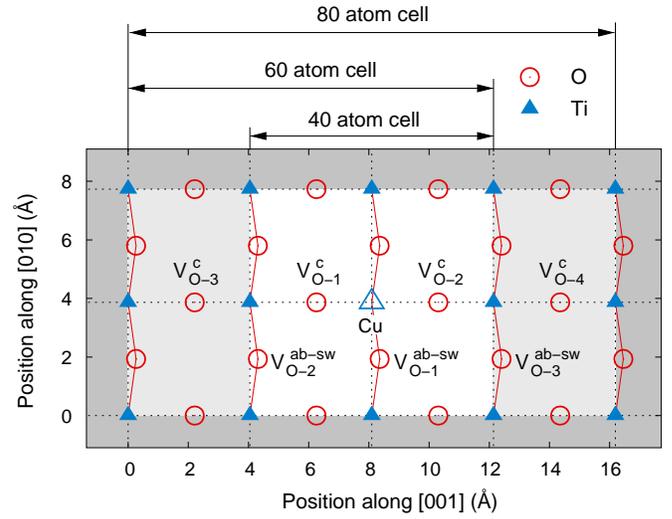}
  \caption{
    (Color online)
    Schematic of different conceivable vacancy-impurity atom
    configurations. The tetragonal perovskite lattice is shown along
    the $[100]$ axis and Pb atoms have been omitted for clarity. The
    different gray shaded areas indicate the size of the different
    supercells employed in the present work.
  }
  \label{fig:confs}
\end{figure}

\begin{figure}
  \centering
\includegraphics[scale=0.7]{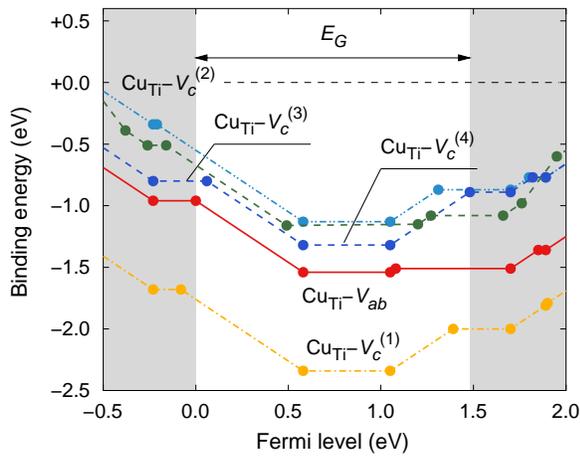}
  \caption{
    (Color online)
    Binding energies for Cu impurities in tetragonal lead
    titanate obtained from the extrapolated formation
    energies given in Tables~\ref{tab:eform_VO} and
    \ref{tab:eform_CuTi}. Negative energies imply a chemical
    driving force for association.
  }
  \label{fig:bind_extrapol_Cu}
\end{figure}

As illustrated in \fig{fig:confs} different configurations for defect
associates of oxygen vacancies with $B$-site impurities are
conceivable. Since the impurity atom breaks the translational symmetry
along the $c$-axis, the degeneracy of the positions above and below
the impurity atom is lifted. In the first neighbor shell one
can distinguish configurations in which the oxygen vacancy is placed
along the $c$-axis either below ($M_{\Ti}-V_c^{(1)}$, shorter $M_{\Ti}-\O$
distance) or above the impurity atom ($M_{\Ti}-V_c^{(2)}$, longer
$M_{\Ti}-\O$ distance), or on one of the four symmetry equivalent
sites within the $ab$-plane ($M_{\Ti}-V_{ab}$). If one considers the
second-nearest neighbor shell, another two configurations involving
$c$-type vacancies are possible ($M_{\Ti}-V_c^{(3)}$,
$M_{\Ti}-V_c^{(4)}$). Further configurations are obtained by placing
the vacancy on second-nearest neighbor $ab$-sites (e.g.,
$M_{\Ti}-V_{ab}^{(2)}$ and $M_{\Ti}-V_{ab}^{(3)}$ in
\fig{fig:confs}). In the present work, we have considered
first and second-nearest neighbor configurations involving $c$-type
vacancies as well as first-nearest neighbor configurations with
$ab$-vacancies. Exploratory calculations were also carried out for
second-nearest neighbor $ab$-type vacancy associates. This possibility
was, however, not pursued further, since the total energies were
significantly larger than for the respective lowest energy
configuration.

The $\Cu_{\Ti}-V_c^{(1)}$ complex, in which the vacancy is located at
the nearest oxygen site along the $c$-axis (compare \fig{fig:confs} and
\fig{fig:dens_CuTi_VO}), is the most stable configuration. The most
stable charge state is $q=0$ as expected if one combines
$V_{\O}^{\mydot\mydot}$ (see \fig{fig:eform_VO}) and
$\Cu_{\Ti}^{''}$. The energy difference with respect to the less
stable complexes ($[\Cu_{\Ti}-V_{ab}]^{\times}$ and
$[\Cu_{\Ti}-V_c^{(3)}]^{\times}$) is rather large
($0.8$-$1.0\,\eV$).
This large energy difference is particular noteworthy since previous
model calculations assumed a much smaller energy difference on the
order of 0.06\,eV (compare Ref.~\onlinecite{ArlNeu88}).

It is very instructive to compare the charge densities for isolated
and complexed copper impurities. As shown in \fig{fig:dens_CuTi}(d),
the spin density for the isolated copper impurity is symmetric with
respect to four-fold rotations about the [001] axis. This symmetry
also applies for the copper-vacancy complex [\fig{fig:dens_CuTi}(h)]
but in addition the spin density pattern contains a (001) mirror plane
through the Cu site. As can be seen by comparison of
Figs.~\ref{fig:dens_CuTi}(a) and (e) this distinction results from the
shift of the Cu atom which in the presence of a vacancy on the nearest
O-site relaxes almost completely into the (001) plane of oxygen atoms
leading to a local pseudo-cubic environment. These observations have
important implications for the interpretation of electron paramagnetic
resonance measurements as discussed in Ref.~\onlinecite{EicErhTra07}.

Finally, one can use the formation energies of isolated oxygen
vacancies, copper impurities and their complexes to derive the binding
energy and its dependence on the Fermi level. As shown in
\fig{fig:bind_extrapol_Cu}, there is a very strong chemical driving
force for association of copper impurities with oxygen vacancies
irrespective of the Fermi levels.
If there are more oxygen vacancies than copper impurities in the
system, all copper impurities will be complexed. In the opposite
scenario all oxygen vacancies would be complexed and the number of
isolated copper impurities would be reduced by the number of metal
impurity-oxygen vacancy associates.

\subsubsection{Iron impurities}

\begingroup
\begin{table}

  \caption{
    Formation energies (LSDA) in units of eV for isolated and
    complexed Fe-impurities on Ti-sites ($\Fe_{\Ti}$) in tetragonal
    lead titanate for a Fermi level at the valence band maximum
    ($\mu_e=0\,\eV$) and metal-rich conditions ($\Delta\mu_{\Pb}
    =\Delta\mu_{\Ti} =\Delta\mu_{\Fe} =0\,\eV$). The finite-size
    scaling extrapolation errors are given in brackets in the last but
    one column.
    $\dnspin$ denotes the difference between the number of electrons
    in spin-up and spin-down states.
  }

  \label{tab:eform_FeTi}

  \centering

  \newcolumntype{C}{>{\centering\centering\arraybackslash}X}
  \newcolumntype{Q}[0]{>{(}d<{)}}
  \newcommand{\spread}[1]{\multicolumn{1}{c}{#1}}
  \newcommand{\spreadd}[1]{\multicolumn{2}{c}{#1}}


  \begin{tabular}{lddddddQd}
    \hline\hline

    \multicolumn{2}{l}{\tbhd{Defect}}
    & \spread{$\boldsymbol{\dnspin}$}
    & \spread{$\boldsymbol{40}$}
    & \spread{$\boldsymbol{60}$}
    & \spread{$\boldsymbol{80}$}
    & \spreadd{\tbhd{extrapolated}}
    \\
    
    \hline
    
    \multicolumn{4}{l}{$\Fe_{\Ti}$} \\
    & -2 & 0 &   8.64 &   8.81 &   8.87 &   9.11 &   0.04 \\
    & -1 & 1 &   7.14 &   7.18 &   7.21 &   7.27 &  <0.01 \\
    & 0  & 2 &   6.26 &   6.23 &   6.23 &   6.19 &   0.01 \\
    & +1 & 3 &   6.15 &   6.12 &   6.10 &   6.05 &  <0.01 \\[6pt]
    
    \multicolumn{4}{l}{$\Fe_{\Ti}-V_{ab}$} \\
    & -2 & 0 &   8.69 &   8.91 &   8.96 &   9.25 &   0.07 \\
    & -1 & 1 &   6.94 &   6.99 &   7.01 &   7.08 &  <0.01 \\
    & 0  & 2 &   5.46 &   5.44 &   5.44 &   5.41 &   0.02 \\
    & +1 & 3 &   4.49 &   4.43 &   4.40 &   4.31 &  <0.01 \\
    & +2 & 4 &   4.18 &   4.14 &   4.09 &   4.01 &   0.04 \\[6pt]

    \multicolumn{4}{l}{$\Fe_{\Ti}-V_c^{(1)}$} \\
    & -2 & 0 &   8.42 &   8.52 &   8.53 &   8.65 &   0.05 \\
    & -1 & 1 &   6.50 &   6.52 &   6.53 &   6.56 &  <0.01 \\
    & 0  & 2 &   5.00 &   4.93 &   4.93 &   4.84 &   0.04 \\
    & +1 & 3 &   3.99 &   3.88 &   3.85 &   3.70 &   0.02 \\
    & +2 & 4 &   3.82 &   3.73 &   3.68 &   3.54 &   0.02 \\[6pt]
    
    \multicolumn{4}{l}{$\Fe_{\Ti}-V_c^{(2)}$} \\
    & -2 & 0 &   8.61 &   9.08 &   9.43 &  10.20 &   0.13 \\
    & -1 & 1 &   6.83 &   7.17 &   7.39 &   7.92 &   0.06 \\
    & 0  & \spread{$^a$}
    &            5.40 &   5.60 &   5.46 &   5.62 &   0.27 \\
    & +1 & \spread{$^a$} 
    &            4.34 &   4.25 &   4.25 &   4.15 &   0.05 \\
    & +2 & 4 &   4.09 &   4.16 &   4.17 &   4.25 &   0.03 \\[6pt]
    
    \multicolumn{4}{l}{$\Fe_{\Ti}-V_c^{(3)}$} \\
    & -2 & 0 &        &   9.19 &   9.11 &   8.86 & \spread{} \\
    & -1 & 0 &        &   6.97 &   6.87 &   6.55 & \spread{} \\
    & 0  & 0 &        &   5.26 &   5.21 &   5.07 & \spread{} \\
    & +1 & 1 &        &   4.32 &   4.23 &   3.96 & \spread{} \\
    & +2 & 2 &        &   4.24 &   4.10 &   3.69 & \spread{} \\[6pt]
    
    \multicolumn{4}{l}{$\Fe_{\Ti}-V_c^{(4)}$} \\
    & -2 & 0 &        &        &   9.42 &        & \spread{} \\
    & -1 & 0 &        &        &   7.21 &        & \spread{} \\
    & 0  & 0 &        &        &   5.55 &        & \spread{} \\
    & +1 & 1 &        &        &   4.54 &        & \spread{} \\
    & +2 & 2 &        &        &   4.27 &        & \spread{} \\[6pt]
    
    \hline\hline
    \\[-8pt]
    
    \multicolumn{9}{p{0.84\columnwidth}}{
      $^a$ The difference between electrons in spin-up and spin-down
      states, $\dnspin$, for the configuration with the lowest energy is
      system size dependent.
    }
  
  \end{tabular}

\end{table}
\endgroup

\begin{figure}
  \centering
\includegraphics[scale=0.7]{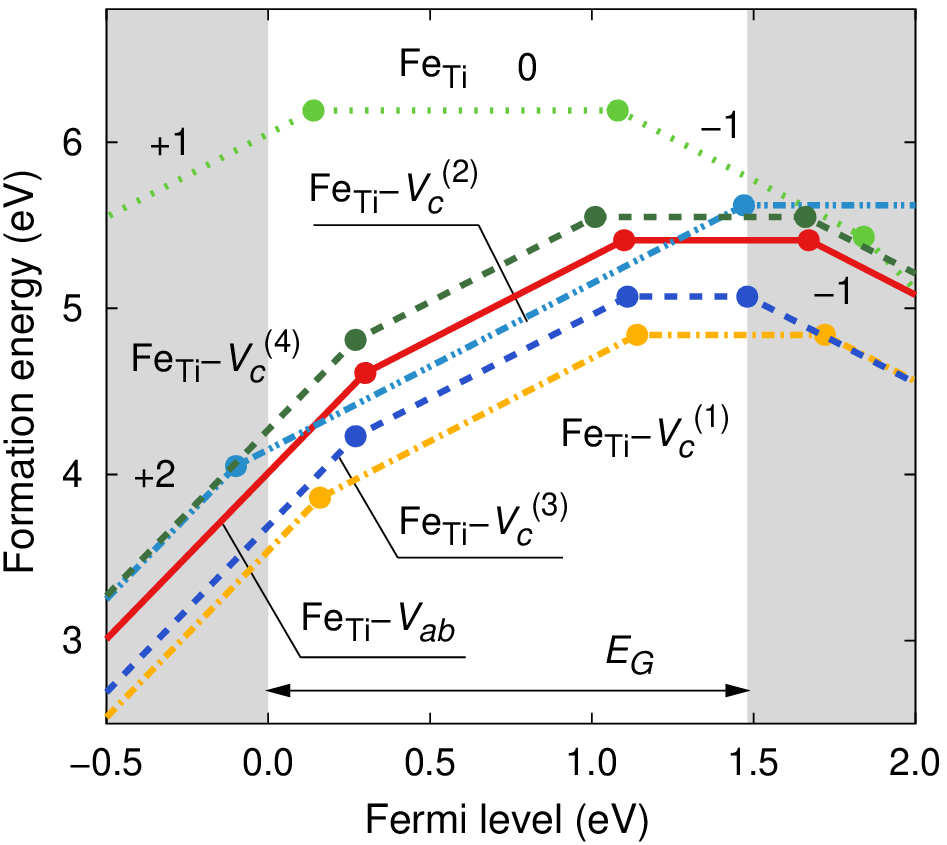}
  \caption{
    (Color online)
    Fermi level dependence of formation energies of free and complexed
    Fe impurities. Parallel lines imply equal charge states as
    indicated in the plot. The white stripe indicates the (calculated)
    band gap while the gray stripes indicate the valence and
    conduction bands, respectively. Extrapolated formation energies
    from \tab{tab:eform_FeTi} were used except for $\Fe_{\Ti}-V_c^{(4)}$
    for which the data obtained with 80-atom cells are shown.
  }
  \label{fig:eform_FeTi}
\end{figure}

\begin{figure}
  \centering
\includegraphics[width=0.9\linewidth]{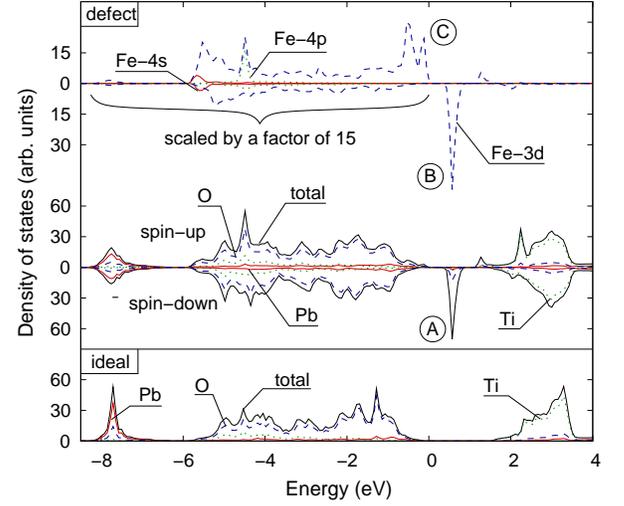}
  \caption{
    (Color online)
    Spin-decomposed total and partial density of states for neutral
    uncomplexed Fe impurities in lead titanate. The partial densities
    of states for the Fe-d orbitals are shown separately in the upper
    part of the figure. The Fe-s and Fe-p states have been omitted
    since in the energy window shown they are practically zero.
    A Gaussian filter with $\sigma=0.05\,\eV$ was employed to smoothen
    the data.
  }
  \label{fig:spin_FeTi}
\end{figure}

\begin{figure}
  \centering
\includegraphics[width=0.96\linewidth]{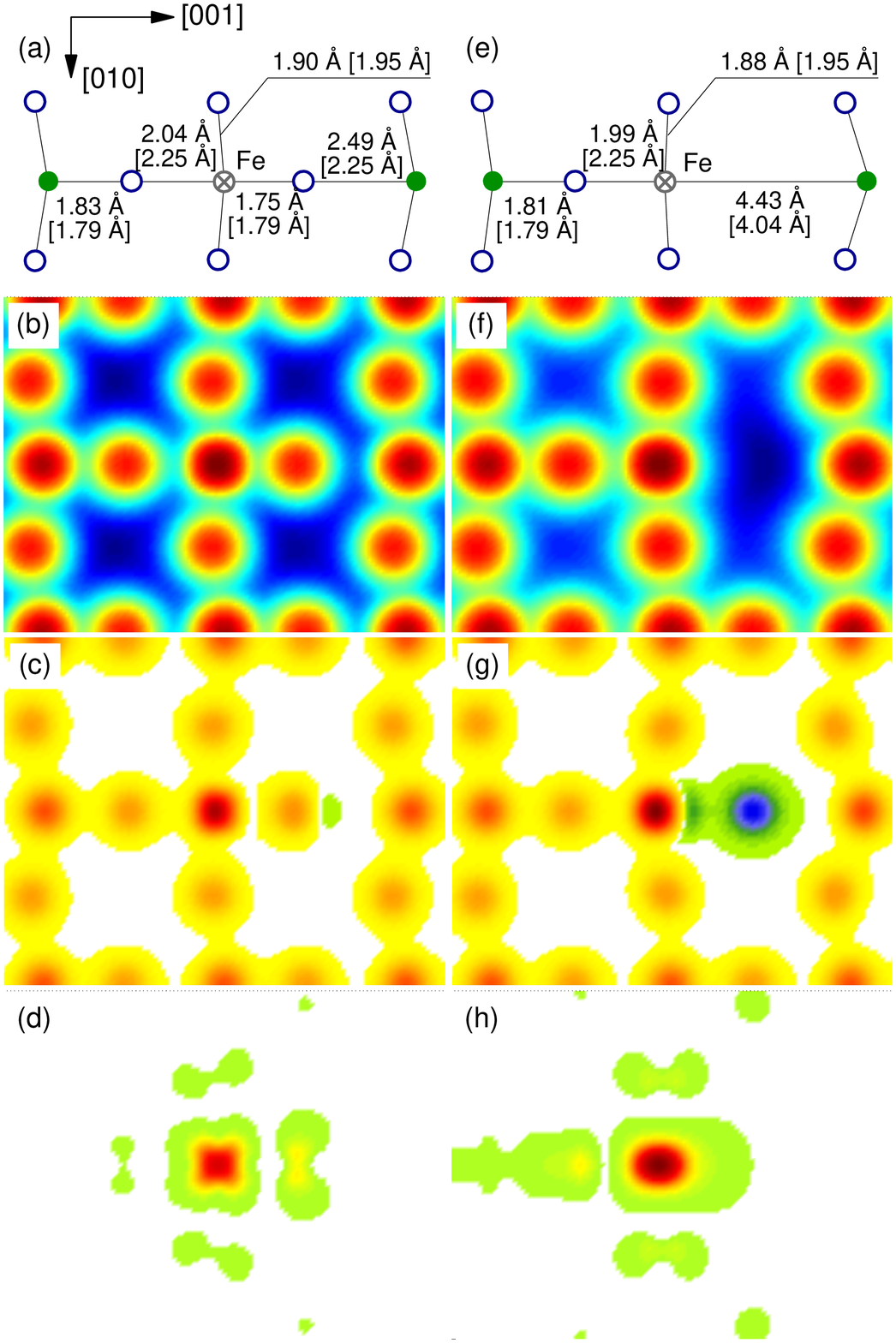}
  \caption{
    (Color online)
    Isolated iron impurity in charge state $q=0$ (a-d) and iron
    impurity-oxygen vacancy complex in charge state $q=+1$ (e-h).
    (a,e) Position of atoms in the (100)-plane containing the defect,
    numbers given in square brackets give the equivalent distances in
    the defect-free crystal;
    (b,f) total charge density, a logarithmic scale has been chosen for
    the charge density in order to enhance the features,
    (c,g) difference between total charge density of defective cell
    and defect-free cell, and
    (d,h) spin densities.
  }
  \label{fig:dens_FeTi}
  \label{fig:dens_FeTi_VO}
\end{figure}

The formation energies for uncomplexed Fe impurities are compiled in
\tab{tab:eform_FeTi} and presented as a function of the Fermi level in
\fig{fig:eform_FeTi}.
The uncomplexed iron impurity occurs predominantly in the neutral
charge state ($q=0$). Transition levels are only present near the band
edges. It is noteworthy that according to \fig{fig:eform_FeTi}
uncomplexed Fe should display ambipolar behavior. Analysis of the
total and partial density of states shows that iron induces a defect
level which for the neutral charge state is located in the middle of
the band gap (feature A in
\fig{fig:spin_FeTi}). The density of states for the two different spin
orientations differ considerably. While one spin-orientation
(spin-down in \fig{fig:spin_FeTi}) gives rise to a level in the band
gap (feature B in \fig{fig:spin_FeTi}), the other one leads to
additional states at the top of the valence band (feature C in
\fig{fig:spin_FeTi}). As in the case of copper, strong hybridization
occurs between the impurity atom 3d-orbitals and the O--2p levels of
the host material. Also, in equivalence to copper Fe--3d levels
do practically not contribute to the conduction band, slightly
diminishing the density of states at the bottom of this band. Due to
the presence of gap states defect induced holes and electrons cannot
be unambiguously identified. In contrast to copper, a simple
correlation between the Kr\"oger-Vink notation for defects and the
charge states, $q$, is, therefore, not possible.

The potentials for the redox reactions $\Fe^{3+} + e^- \rightarrow
\Fe^{++}$ and $\Fe^{++} + 2 e^- \rightarrow \Fe^0$ have been measured in
aqueous solution to be 0.77\,eV and $-0.41\,\eV$, respectively. Based on
these values and using the data for the electron affinity and the band
gap cited above, one would expect transition levels at 1.6\,eV
($\Fe^{3+}/\Fe^{++}$, i.e. $\Fe_{\Ti}^{'}/\Fe_{\Ti}^{''}$) and 2.8\,eV
($\Fe^{++}/\Fe^{0}$, i.e. $\Fe_{\Ti}^{''}/\Fe_{\Ti}^{4'}$) with
respect to the experimental valence band maximum.
In contrast, the present calculations indicate transitions about
0.15\,eV above the VBM and 0.39\,eV below the CBM. The analogy with
the behavior of ions in solution, which worked reasonably in the case
of copper, thus appears to fail for iron.

\subsubsection{Iron-vacancy complexes}
\label{sect:FeVO}

The iron-vacancy complexes which are analogous to the copper-vacancy
associates described in \sect{sect:CuVO} and \fig{fig:confs}. In
agreement with earlier calculations \cite{MesEicKlo05} the
configuration with the shortest distance between impurity center and
oxygen vacancy ($\Fe_{\Ti}-V_c^{(1)}$) is the most stable
(\fig{fig:eform_FeTi}). The energetic ordering of the remaining
configurations is, however, somewhat different from the case of
copper. This observation has implications for the migration of oxygen
vacancies and is discussed in the context of point defect models in
\sect{sect:discussion} and Ref.~\onlinecite{ErhTraAlb07}.

\begin{figure}
  \centering
\includegraphics[scale=0.7]{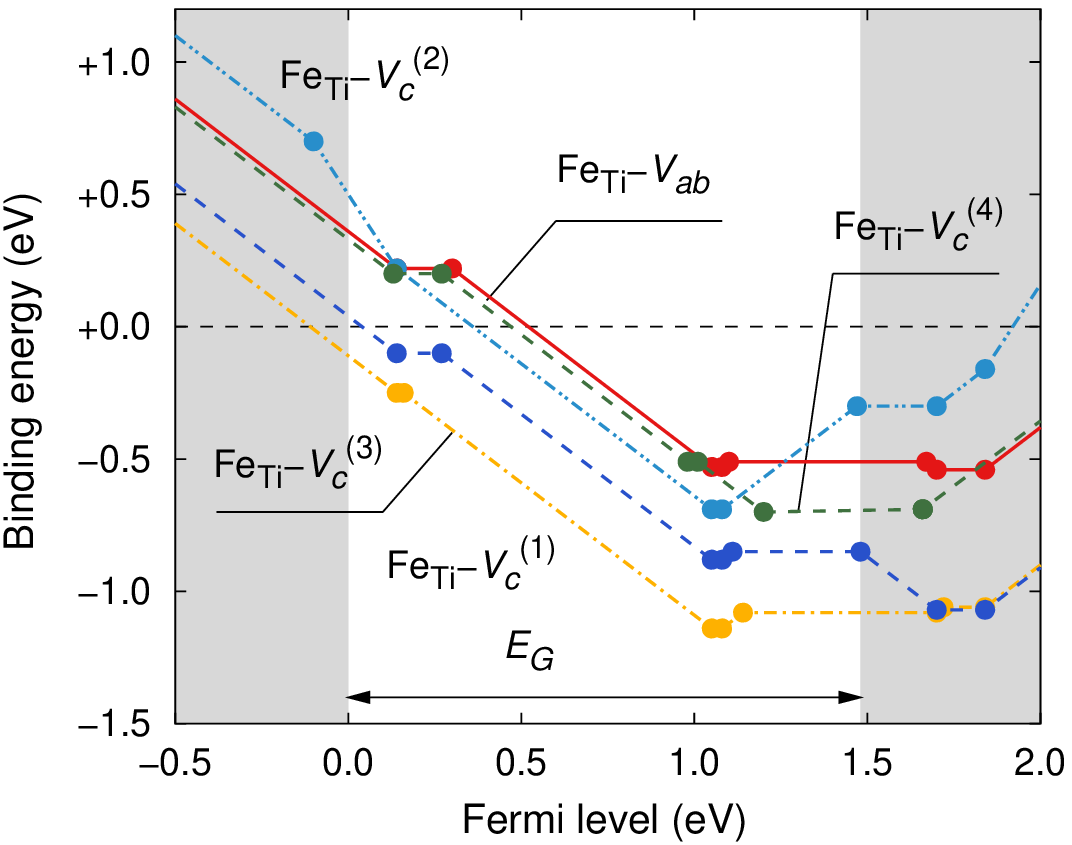}
  \caption{
    (Color online)
    Binding energies as defined by equation \eq{eq:ebind} for Fe
    impurities in tetragonal lead titanate obtained from the
    extrapolated formation energies given in Tables~\ref{tab:eform_VO}
    and \ref{tab:eform_FeTi}. Negative energies imply a chemical
    driving force for association.
  }
  \label{fig:bind_extrapol_Fe}
\end{figure}

The binding energies for Fe impurities shown in
\fig{fig:bind_extrapol_Fe} are in general smaller than for
copper and for Fermi levels very close to the valence band the binding
energy can even become positive. Nonetheless for most conditions there
is a rather strong ($\gg k_B T$) driving force for association.

Analysis of the spin density shows that for both isolated and
associated iron impurities unpaired electrons are localized in the
d-electron states at the iron site (\tab{tab:eform_FeTi}). It
furthermore reveals that in both cases the spin density pattern is
symmetric with respect to fourfold rotations about the tetragonal
axis. Unlike in the case of copper-vacancy associates the (001) plane
is, however, not a mirror plane.

\section{Discussion}
\label{sect:discussion}

\begin{figure*}
  \centering
\includegraphics[width=\linewidth]{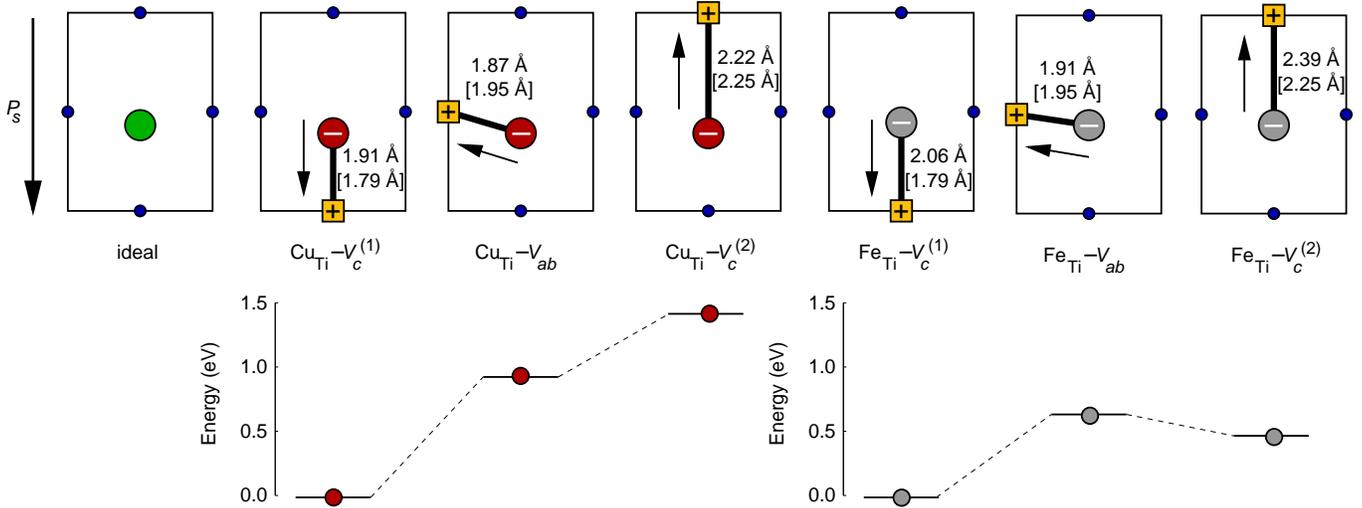}
  \caption{
    (Color online)
    Summary of key results for the most important charge states of
    nearest neighbor Cu/Fe-vacancy complexes
    ($[\Cu_{\Ti}-V_{\O}]^{\times}$, $[\Fe_{\Ti}-V_{\O}]^{\mydot}$).
    The large arrow on the far left indicates the direction
    of the spontaneous polarization with respect to the ideal unit
    cell. The short arrows indicate the direction of the defect
    induced dipole moments. Distances are given between the impurity
    atom and the position of the vacancy (as defined by the positions
    of the neighboring oxygen planes. Values in square brackets refer
    to the ideal (undistorted) lattice. The lower panel visualizes the
    formation energy differences.
  }
  \label{fig:schematic}
\end{figure*}

The results obtained in this study need to be discussed in the context
of degradation phenomena in ferroelectric materials. Two distinct
processes have been widely described:
(1)
The gradual degradation of ferroelectric properties (usually over
extended periods of time) in the absence of electric fields is termed
``aging''. \cite{LamJon86} This process is commonly
attributed to an increasing restriction of domain-wall motion with
time \cite{Arl87} and typically accompanied by a shift of the
ferroelectric hysteresis loop along the electric field
axis. \cite{CarHar78, Tak82} The shift is known as the {\it internal
  bias field} and it has been proposed to be related to the relaxation
of defect dipoles. \cite{NeuArl87, ArlNeu88} In general,
acceptor-doped ferroelectric ceramics with high concentrations of
oxygen vacancies are particularly prone to aging.
(2)
The degradation of ferroelectric properties in the presence of a
(usually oscillating) electric field is termed electric
``fatigue''. There are various contributions to this phenomenon and
possible explanations involve defects of all kinds.

The most widely known model for aging \cite{NeuArl87, ArlNeu88}
relates the appearance of internal bias fields to the existence and
gradual reorientation of impurity-oxygen vacancy defect
dipoles. Similar models have been discussed in the literature
\cite{WarDimPik95, WarPikVan96, Ren04}
Experimentally, an increased internal bias field has been observed for
aged lead zirconate-titanate compounds and was attributed to defect
dipole alignment leading to a clamping of domain
walls.\cite{RobZadArl92}

The availability of microscopic information on the energy landscape
for defect formation and migration is the key ingredient for the these
models. The present work provides a data base for these model and
allow to verify their basic assumptions.

The key information regarding the orientation and the magnitude of the
defect dipoles as well as their energetic ordering is summarized in
\fig{fig:schematic}. This picture needs to be compared to the energy
landscape proposed by Arlt and Neumann \cite{NeuArl87, ArlNeu88} on
the basis of electrostatic arguments (Figs.~4 in
Refs.~\onlinecite{NeuArl87} and \onlinecite{ArlNeu88}). The latter
analysis was carried out for barium titanate and involved several
crucial assumptions with respect to e.g., the magnitude of the local
dipole moment or the dielectric constant in the vicinity of the defect
center. In agreement with the conclusions of the Arlt-Neumann model, the
quantum-mechanical calculations show that for the energetically most
stable configuration ($M_{\Ti}-V_c^{(1)}$) the defect induced
dipole is oriented parallel to the overall spontaneous polarization,
$P_s$ (compare \fig{fig:schematic}). Thus in equilibrium (i.e., a
state which would be achieved after ``infinite'' aging) the
polarization locally increases due the formation and alignment of local
defect dipoles. In contrast, the defect dipole of the alternative
$c$-axis configuration ($M_{\Ti}-V_c^{(2)}$) causes a local reduction of
the polarization.

In Ref.~\onlinecite{ArlNeu88} two estimates for the energy differences
between different dipole orientations are derived. The first one
treats the defect as a sphere with a defect induced polarization
embedded in a ferroelectric matrix. It yields values on the order of
0.5 to 1.2\,eV (compare equations (8a-8c) in
Ref.~\onlinecite{ArlNeu88}) which are of a similar magnitude as the
values obtained from the present calculations. The second estimate is
based on the model of an ideal point dipole and yields much smaller
energy differences on the order of 0.03 to 0.06\,eV (equation (9) in
Ref.~\onlinecite{ArlNeu88}). It is important to realize that the
modeling of aging carried out in the same paper was actually
based on the latter values.

In fact the energy differences obtained in the present work are
actually on the order of magnitude of the barriers for the migration
of free oxygen vacancies in this system (0.8-1.0\,eV, see
e.g. Ref.~\onlinecite{Par03}). In the vicinity of impurities the
energy surface, which describes the barriers for the transformation of
different defect dipole configurations into each other, is therefore
severely distorted.\cite{ErhTraAlb07} This effect is important for
the behavior of defect dipoles in the presence of oscillating electric
fields, which pertains to the fatigue of the material.

Finally, \fig{fig:schematic} highlights a noteworthy difference
between Cu and Fe. While for Cu-complexes the energy difference
between $[\Cu_{\Ti}-V_c^{(1)}]^{\times}$ and
$[\Cu_{\Ti}-V_c^{(2)}]^{\times}$ is as large as $1.21\,\eV$, for
$[\Fe_{\Ti}-V_c^{(1)}]^{\mydot}$ and $[\Fe_{\Ti}-V_c^{(2)}]^{\mydot}$
the difference is just 0.45\,eV. This difference has implications for
the redistribution of the defect dipoles in the presence of an
oscillating electric field and thus for the process of fatigue. A
detailed account of these behavior will be published
elsewhere.\cite{ErhTraAlb07}

\section{Conclusions}

This paper presents a comprehensive and detailed investigation of Fe
and Cu dopants/impurities in tetragonal lead titanate. Both metal ions
are found to be very effective traps for oxygen vacancies. In the most
stable configurations the oxygen vacancy is located on a $c$-site such
that the $M_{\Ti}-V_{\O}$ distance is minimal, and the defect dipole
is aligned parallel the spontaneous polarization. This result is in
qualitative agreement with an electrostatic analysis of the defect
induced dipole field.
An alternative configuration ($M_{\Ti}-V_c^{(2)}$) in which the dipole
is oriented anti-parallel with respect to the spontaneous polarization
is somewhat higher energy but is anticipated to be of importance for
the fatigue of lead titanate based ferroelectrics.

Since all relevant defects considered in this investigation carry
unpaired electrons, they are visible for electron spectroscopic
methods. In fact, by means of analyzing the spin density patterns for
different defects it was possible to interpret recent electron
paramagnetic resonance measurements and to resolve an apparent
disagreement between experiment and calculations.\cite{EicErhTra07} The
results obtained in this study represent important information on the
energy landscape for energy formation and association and provide the
basis for the improvement of defect models for the degradation of
ferroelectric materials.

\begin{acknowledgments}
This project was funded by the \textit{Sonderforschungsbereich 595}
``Fatigue in functional materials'' of the \textit{Deutsche
  Forschungsgemeinschaft}. Fruitful discussions with N. Balke
(University of California, Berkeley) and S. Gottschalk
(Forschungszentrum Karlsruhe) are gratefully acknowledged.
\end{acknowledgments}


\begin{thebibliography}{49}
\expandafter\ifx\csname natexlab\endcsname\relax\def\natexlab#1{#1}\fi
\expandafter\ifx\csname bibnamefont\endcsname\relax
  \def\bibnamefont#1{#1}\fi
\expandafter\ifx\csname bibfnamefont\endcsname\relax
  \def\bibfnamefont#1{#1}\fi
\expandafter\ifx\csname citenamefont\endcsname\relax
  \def\citenamefont#1{#1}\fi
\expandafter\ifx\csname url\endcsname\relax
  \def\url#1{\texttt{#1}}\fi
\expandafter\ifx\csname urlprefix\endcsname\relax\def\urlprefix{URL }\fi
\providecommand{\bibinfo}[2]{#2}
\providecommand{\eprint}[2][]{\url{#2}}

\bibitem[{\citenamefont{Warren et~al.}(1995{\natexlab{a}})\citenamefont{Warren,
  Tuttle, and Dimos}}]{WarTutDim95}
\bibinfo{author}{\bibfnamefont{W.~L.} \bibnamefont{Warren}},
  \bibinfo{author}{\bibfnamefont{B.~A.} \bibnamefont{Tuttle}},
  \bibnamefont{and} \bibinfo{author}{\bibfnamefont{D.}~\bibnamefont{Dimos}},
  \bibinfo{journal}{Appl. Phys. Lett.} \textbf{\bibinfo{volume}{67}},
  \bibinfo{pages}{1426} (\bibinfo{year}{1995}{\natexlab{a}}).

\bibitem[{\citenamefont{Warren et~al.}(1995{\natexlab{b}})\citenamefont{Warren,
  Dimos, Pike, Vanheusden, and Ramesh}}]{WarDimPik95}
\bibinfo{author}{\bibfnamefont{W.~L.} \bibnamefont{Warren}},
  \bibinfo{author}{\bibfnamefont{D.}~\bibnamefont{Dimos}},
  \bibinfo{author}{\bibfnamefont{G.~E.} \bibnamefont{Pike}},
  \bibinfo{author}{\bibfnamefont{K.}~\bibnamefont{Vanheusden}},
  \bibnamefont{and} \bibinfo{author}{\bibfnamefont{R.}~\bibnamefont{Ramesh}},
  \bibinfo{journal}{Appl. Phys. Lett.} \textbf{\bibinfo{volume}{67}},
  \bibinfo{pages}{1689} (\bibinfo{year}{1995}{\natexlab{b}}).

\bibitem[{\citenamefont{Me\v{s}tri\'c et~al.}(2004)\citenamefont{Me\v{s}tri\'c,
  Eichel, Dinse, Ozarowski, van Tol, and Brunel}}]{MesEicDin04}
\bibinfo{author}{\bibfnamefont{H.}~\bibnamefont{Me\v{s}tri\'c}},
  \bibinfo{author}{\bibfnamefont{R.-A.} \bibnamefont{Eichel}},
  \bibinfo{author}{\bibfnamefont{K.-P.} \bibnamefont{Dinse}},
  \bibinfo{author}{\bibfnamefont{A.}~\bibnamefont{Ozarowski}},
  \bibinfo{author}{\bibfnamefont{J.}~\bibnamefont{van Tol}}, \bibnamefont{and}
  \bibinfo{author}{\bibfnamefont{L.}~\bibnamefont{Brunel}},
  \bibinfo{journal}{J. Appl. Phys.} \textbf{\bibinfo{volume}{96}},
  \bibinfo{pages}{7440} (\bibinfo{year}{2004}).

\bibitem[{\citenamefont{Me\v{s}tri\'{c}
  et~al.}(2005)\citenamefont{Me\v{s}tri\'{c}, Eichel, Kloss, Dinse, Laubach,
  Laubach, Schmidt, Sch\"onau, Knapp, and Ehrenberg}}]{MesEicKlo05}
\bibinfo{author}{\bibfnamefont{H.}~\bibnamefont{Me\v{s}tri\'{c}}},
  \bibinfo{author}{\bibfnamefont{R.-A.} \bibnamefont{Eichel}},
  \bibinfo{author}{\bibfnamefont{T.}~\bibnamefont{Kloss}},
  \bibinfo{author}{\bibfnamefont{K.-P.} \bibnamefont{Dinse}},
  \bibinfo{author}{\bibfnamefont{S.}~\bibnamefont{Laubach}},
  \bibinfo{author}{\bibfnamefont{S.}~\bibnamefont{Laubach}},
  \bibinfo{author}{\bibfnamefont{P.~C.} \bibnamefont{Schmidt}},
  \bibinfo{author}{\bibfnamefont{K.~A.} \bibnamefont{Sch\"onau}},
  \bibinfo{author}{\bibfnamefont{M.}~\bibnamefont{Knapp}}, \bibnamefont{and}
  \bibinfo{author}{\bibfnamefont{H.}~\bibnamefont{Ehrenberg}},
  \bibinfo{journal}{Phys. Rev. B} \textbf{\bibinfo{volume}{71}},
  \bibinfo{pages}{134109} (\bibinfo{year}{2005}).

\bibitem[{\citenamefont{P\"oykk\"o and Chadi}(1999)}]{PoyCha99a}
\bibinfo{author}{\bibfnamefont{S.}~\bibnamefont{P\"oykk\"o}} \bibnamefont{and}
  \bibinfo{author}{\bibfnamefont{D.~J.} \bibnamefont{Chadi}},
  \bibinfo{journal}{Phys. Rev. Lett.} \textbf{\bibinfo{volume}{83}},
  \bibinfo{pages}{1231} (\bibinfo{year}{1999}).

\bibitem[{\citenamefont{Neumann and Arlt}(1987)}]{NeuArl87}
\bibinfo{author}{\bibfnamefont{H.}~\bibnamefont{Neumann}} \bibnamefont{and}
  \bibinfo{author}{\bibfnamefont{G.}~\bibnamefont{Arlt}},
  \bibinfo{journal}{Ferroelectrics} \textbf{\bibinfo{volume}{76}},
  \bibinfo{pages}{303} (\bibinfo{year}{1987}).

\bibitem[{\citenamefont{Arlt and Neumann}(1988)}]{ArlNeu88}
\bibinfo{author}{\bibfnamefont{G.}~\bibnamefont{Arlt}} \bibnamefont{and}
  \bibinfo{author}{\bibfnamefont{H.}~\bibnamefont{Neumann}},
  \bibinfo{journal}{Ferroelectrics} \textbf{\bibinfo{volume}{87}},
  \bibinfo{pages}{109} (\bibinfo{year}{1988}).

\bibitem[{\citenamefont{Park and Chadi}(1998)}]{ParCha98}
\bibinfo{author}{\bibfnamefont{C.~H.} \bibnamefont{Park}} \bibnamefont{and}
  \bibinfo{author}{\bibfnamefont{D.~J.} \bibnamefont{Chadi}},
  \bibinfo{journal}{Phys. Rev. B} \textbf{\bibinfo{volume}{57}},
  \bibinfo{pages}{R13961} (\bibinfo{year}{1998}).

\bibitem[{\citenamefont{P\"oykk\"o and Chadi}(2000{\natexlab{a}})}]{PoyCha00a}
\bibinfo{author}{\bibfnamefont{S.}~\bibnamefont{P\"oykk\"o}} \bibnamefont{and}
  \bibinfo{author}{\bibfnamefont{D.~J.} \bibnamefont{Chadi}},
  \bibinfo{journal}{Appl. Phys. Lett.} \textbf{\bibinfo{volume}{76}},
  \bibinfo{pages}{499} (\bibinfo{year}{2000}{\natexlab{a}}).

\bibitem[{\citenamefont{Cockayne and Burton}(2004)}]{CocBur04}
\bibinfo{author}{\bibfnamefont{E.}~\bibnamefont{Cockayne}} \bibnamefont{and}
  \bibinfo{author}{\bibfnamefont{B.~P.} \bibnamefont{Burton}},
  \bibinfo{journal}{Phys. Rev. B} \textbf{\bibinfo{volume}{69}},
  \bibinfo{pages}{144116} (\bibinfo{year}{2004}).

\bibitem[{\citenamefont{Park}(2003)}]{Par03}
\bibinfo{author}{\bibfnamefont{C.~H.} \bibnamefont{Park}}, \bibinfo{journal}{J.
  Korean Phys. Soc.} \textbf{\bibinfo{volume}{42}}, \bibinfo{pages}{S1420 }
  (\bibinfo{year}{2003}).

\bibitem[{\citenamefont{He and Vanderbilt}(2003)}]{HeVan03}
\bibinfo{author}{\bibfnamefont{L.}~\bibnamefont{He}} \bibnamefont{and}
  \bibinfo{author}{\bibfnamefont{D.}~\bibnamefont{Vanderbilt}},
  \bibinfo{journal}{Phys. Rev. B} \textbf{\bibinfo{volume}{68}},
  \bibinfo{pages}{134103} (\bibinfo{year}{2003}).

\bibitem[{\citenamefont{Zhang et~al.}(2006)\citenamefont{Zhang, Wu, Lu, and
  Shu}}]{ZhaWuLu06}
\bibinfo{author}{\bibfnamefont{Z.}~\bibnamefont{Zhang}},
  \bibinfo{author}{\bibfnamefont{P.}~\bibnamefont{Wu}},
  \bibinfo{author}{\bibfnamefont{L.}~\bibnamefont{Lu}}, \bibnamefont{and}
  \bibinfo{author}{\bibfnamefont{C.}~\bibnamefont{Shu}},
  \bibinfo{journal}{Appl. Phys. Lett.} \textbf{\bibinfo{volume}{88}},
  \bibinfo{pages}{142902} (\bibinfo{year}{2006}).

\bibitem[{\citenamefont{P\"oykk\"o and Chadi}(2000{\natexlab{b}})}]{PoyCha00b}
\bibinfo{author}{\bibfnamefont{S.}~\bibnamefont{P\"oykk\"o}} \bibnamefont{and}
  \bibinfo{author}{\bibfnamefont{D.~J.} \bibnamefont{Chadi}},
  \bibinfo{journal}{J. Phys. Chem. Solids} \textbf{\bibinfo{volume}{61}},
  \bibinfo{pages}{291} (\bibinfo{year}{2000}{\natexlab{b}}).

\bibitem[]{EicErhTra07}
\bibinfo{author}{\bibfnamefont{R.-A.} \bibnamefont{Eichel}},
  \bibinfo{author}{\bibfnamefont{P.}~\bibnamefont{Erhart}},
  \bibinfo{author}{\bibfnamefont{P.}~\bibnamefont{Tr\"askelin}},
  \bibinfo{author}{\bibfnamefont{K.}~\bibnamefont{Albe}},
  \bibinfo{author}{\bibfnamefont{H.}~\bibnamefont{Kungl}},
  and \bibinfo{author}{\bibfnamefont{M. J.}~\bibnamefont{Hoffmann}},
  \bibinfo{journal}{Phys. Rev. Lett.} \textbf{\bibinfo{volume}{100}},
  \bibinfo{pages}{095504} (\bibinfo{year}{2008}).

\bibitem[{\citenamefont{Kresse and Hafner}(1993)}]{KreHaf93}
\bibinfo{author}{\bibfnamefont{G.}~\bibnamefont{Kresse}} \bibnamefont{and}
  \bibinfo{author}{\bibfnamefont{J.}~\bibnamefont{Hafner}},
  \bibinfo{journal}{Phys. Rev. B} \textbf{\bibinfo{volume}{47}},
  \bibinfo{pages}{558} (\bibinfo{year}{1993}).

\bibitem[{\citenamefont{Kresse and Hafner}(1994)}]{KreHaf94}
\bibinfo{author}{\bibfnamefont{G.}~\bibnamefont{Kresse}} \bibnamefont{and}
  \bibinfo{author}{\bibfnamefont{J.}~\bibnamefont{Hafner}},
  \bibinfo{journal}{Phys. Rev. B} \textbf{\bibinfo{volume}{49}},
  \bibinfo{pages}{14251} (\bibinfo{year}{1994}).

\bibitem[{\citenamefont{Kresse and
  Furthm\"uller}(1996{\natexlab{a}})}]{KreFur96a}
\bibinfo{author}{\bibfnamefont{G.}~\bibnamefont{Kresse}} \bibnamefont{and}
  \bibinfo{author}{\bibfnamefont{J.}~\bibnamefont{Furthm\"uller}},
  \bibinfo{journal}{Phys. Rev. B} \textbf{\bibinfo{volume}{54}},
  \bibinfo{pages}{11169} (\bibinfo{year}{1996}{\natexlab{a}}).

\bibitem[{\citenamefont{Kresse and
  Furthm\"uller}(1996{\natexlab{b}})}]{KreFur96b}
\bibinfo{author}{\bibfnamefont{G.}~\bibnamefont{Kresse}} \bibnamefont{and}
  \bibinfo{author}{\bibfnamefont{J.}~\bibnamefont{Furthm\"uller}},
  \bibinfo{journal}{Comput. Mater. Sci.} \textbf{\bibinfo{volume}{6}},
  \bibinfo{pages}{15} (\bibinfo{year}{1996}{\natexlab{b}}).

\bibitem[{\citenamefont{Bl\"ochl}(1994)}]{Blo94}
\bibinfo{author}{\bibfnamefont{P.~E.} \bibnamefont{Bl\"ochl}},
  \bibinfo{journal}{Phys. Rev. B} \textbf{\bibinfo{volume}{50}},
  \bibinfo{pages}{17953} (\bibinfo{year}{1994}).

\bibitem[{\citenamefont{Kresse and Joubert}(1999)}]{KreJou99}
\bibinfo{author}{\bibfnamefont{G.}~\bibnamefont{Kresse}} \bibnamefont{and}
  \bibinfo{author}{\bibfnamefont{D.}~\bibnamefont{Joubert}},
  \bibinfo{journal}{Phys. Rev. B} \textbf{\bibinfo{volume}{59}},
  \bibinfo{pages}{1758 } (\bibinfo{year}{1999}).

\bibitem[{\citenamefont{Monkhorst and Pack}(1976)}]{MonPac76}
\bibinfo{author}{\bibfnamefont{H.~J.} \bibnamefont{Monkhorst}}
  \bibnamefont{and} \bibinfo{author}{\bibfnamefont{J.~D.} \bibnamefont{Pack}},
  \bibinfo{journal}{Phys. Rev. B} \textbf{\bibinfo{volume}{13}},
  \bibinfo{pages}{5188} (\bibinfo{year}{1976}).

\bibitem[{\citenamefont{Persson et~al.}(2005)\citenamefont{Persson, Zhao, Lany,
  and Zunger}}]{PerZhaLan05}
\bibinfo{author}{\bibfnamefont{C.}~\bibnamefont{Persson}},
  \bibinfo{author}{\bibfnamefont{Y.-J.} \bibnamefont{Zhao}},
  \bibinfo{author}{\bibfnamefont{S.}~\bibnamefont{Lany}}, \bibnamefont{and}
  \bibinfo{author}{\bibfnamefont{A.}~\bibnamefont{Zunger}},
  \bibinfo{journal}{Phys. Rev. B} \textbf{\bibinfo{volume}{72}},
  \bibinfo{pages}{035211} (\bibinfo{year}{2005}).

\bibitem[{\citenamefont{Janotti and Van~de Walle}(2006)}]{JanVan06}
\bibinfo{author}{\bibfnamefont{A.}~\bibnamefont{Janotti}} \bibnamefont{and}
  \bibinfo{author}{\bibfnamefont{C.~G.} \bibnamefont{Van~de Walle}},
  \bibinfo{journal}{J. Cryst. Growth} \textbf{\bibinfo{volume}{287}},
  \bibinfo{pages}{58} (\bibinfo{year}{2006}).

\bibitem[{\citenamefont{Erhart and Albe}()}]{ErhAlb07a}
\bibinfo{author}{\bibfnamefont{P.}~\bibnamefont{Erhart}} \bibnamefont{and}
  \bibinfo{author}{\bibfnamefont{K.}~\bibnamefont{Albe}},
  \bibinfo{note}{J. Appl. Phys.} \textbf{\bibinfo{volume}{102}},
  \bibinfo{pages}{084111} (\bibinfo{year}{2007}).

\bibitem[{\citenamefont{Lento et~al.}(2002)\citenamefont{Lento, Mozos, and
  Nieminen}}]{LenMozNie02}
\bibinfo{author}{\bibfnamefont{J.}~\bibnamefont{Lento}},
  \bibinfo{author}{\bibfnamefont{J.-L.} \bibnamefont{Mozos}}, \bibnamefont{and}
  \bibinfo{author}{\bibfnamefont{R.~M.} \bibnamefont{Nieminen}},
  \bibinfo{journal}{J. Phys.: Condens. Matter} \textbf{\bibinfo{volume}{14}},
  \bibinfo{pages}{2637} (\bibinfo{year}{2002}).

\bibitem[{\citenamefont{Erhart et~al.}(2006)\citenamefont{Erhart, Albe, and
  Klein}}]{ErhAlbKle06}
\bibinfo{author}{\bibfnamefont{P.}~\bibnamefont{Erhart}},
  \bibinfo{author}{\bibfnamefont{K.}~\bibnamefont{Albe}}, \bibnamefont{and}
  \bibinfo{author}{\bibfnamefont{A.}~\bibnamefont{Klein}},
  \bibinfo{journal}{Phys. Rev. B} \textbf{\bibinfo{volume}{73}},
  \bibinfo{pages}{205203} (\bibinfo{year}{2006}).

\bibitem[{\citenamefont{Makov and Payne}(1995)}]{MakPay95}
\bibinfo{author}{\bibfnamefont{G.}~\bibnamefont{Makov}} \bibnamefont{and}
  \bibinfo{author}{\bibfnamefont{M.~C.} \bibnamefont{Payne}},
  \bibinfo{journal}{Phys. Rev. B} \textbf{\bibinfo{volume}{51}},
  \bibinfo{pages}{4014} (\bibinfo{year}{1995}).

\bibitem[{\citenamefont{Qian et~al.}(1988)\citenamefont{Qian, Martin, and
  Chadi}}]{QiaMarCha88}
\bibinfo{author}{\bibfnamefont{G.-X.} \bibnamefont{Qian}},
  \bibinfo{author}{\bibfnamefont{R.~M.} \bibnamefont{Martin}},
  \bibnamefont{and} \bibinfo{author}{\bibfnamefont{D.~J.} \bibnamefont{Chadi}},
  \bibinfo{journal}{Phys. Rev. B} \textbf{\bibinfo{volume}{38}},
  \bibinfo{pages}{7649} (\bibinfo{year}{1988}).

\bibitem[{\citenamefont{Zhang and Northrup}(1991)}]{ZhaNor91}
\bibinfo{author}{\bibfnamefont{S.~B.} \bibnamefont{Zhang}} \bibnamefont{and}
  \bibinfo{author}{\bibfnamefont{J.~E.} \bibnamefont{Northrup}},
  \bibinfo{journal}{Phys. Rev. Lett.} \textbf{\bibinfo{volume}{67}},
  \bibinfo{pages}{2339} (\bibinfo{year}{1991}).

\bibitem[{\citenamefont{Erhart et~al.}()\citenamefont{Erhart, Tr\"askelin, and
  Albe}}]{ErhTraAlb07}
\bibinfo{author}{\bibfnamefont{P.}~\bibnamefont{Erhart}},
  \bibinfo{author}{\bibfnamefont{P.}~\bibnamefont{Tr\"askelin}},
  \bibnamefont{and} \bibinfo{author}{\bibfnamefont{K.}~\bibnamefont{Albe}},
  \bibinfo{note}{unpublished}.

\bibitem[{\citenamefont{Eichel et~al.}(2004)\citenamefont{Eichel, Kungl, and
  Hoffmann}}]{EicKunHof04}
\bibinfo{author}{\bibfnamefont{R.~A.} \bibnamefont{Eichel}},
  \bibinfo{author}{\bibfnamefont{H.}~\bibnamefont{Kungl}}, \bibnamefont{and}
  \bibinfo{author}{\bibfnamefont{M.~J.} \bibnamefont{Hoffmann}},
  \bibinfo{journal}{J. Appl. Phys.} \textbf{\bibinfo{volume}{95}},
  \bibinfo{pages}{8092} (\bibinfo{year}{2004}).

\bibitem[{\citenamefont{Eichel et~al.}(2005{\natexlab{a}})\citenamefont{Eichel,
  Dinse, Kungl, Hoffmann, Ozarowski, van Tol, and Brunel}}]{EicDinKun05}
\bibinfo{author}{\bibfnamefont{R.~A.} \bibnamefont{Eichel}},
  \bibinfo{author}{\bibfnamefont{K.~P.} \bibnamefont{Dinse}},
  \bibinfo{author}{\bibfnamefont{H.}~\bibnamefont{Kungl}},
  \bibinfo{author}{\bibfnamefont{M.~J.} \bibnamefont{Hoffmann}},
  \bibinfo{author}{\bibfnamefont{A.}~\bibnamefont{Ozarowski}},
  \bibinfo{author}{\bibfnamefont{J.}~\bibnamefont{van Tol}}, \bibnamefont{and}
  \bibinfo{author}{\bibfnamefont{L.~C.} \bibnamefont{Brunel}},
  \bibinfo{journal}{Appl. Phys. A} \textbf{\bibinfo{volume}{80}},
  \bibinfo{pages}{51} (\bibinfo{year}{2005}{\natexlab{a}}).

\bibitem[{\citenamefont{Eichel et~al.}(2005{\natexlab{b}})\citenamefont{Eichel,
  Me\v{s}tri\'{c}, Dinse, Ozarowski, van Tol, Brunel, Kungl, and
  Hoffmann}}]{EicMesDin05}
\bibinfo{author}{\bibfnamefont{R.~A.} \bibnamefont{Eichel}},
  \bibinfo{author}{\bibfnamefont{H.}~\bibnamefont{Me\v{s}tri\'{c}}},
  \bibinfo{author}{\bibfnamefont{K.~P.} \bibnamefont{Dinse}},
  \bibinfo{author}{\bibfnamefont{A.}~\bibnamefont{Ozarowski}},
  \bibinfo{author}{\bibfnamefont{J.}~\bibnamefont{van Tol}},
  \bibinfo{author}{\bibfnamefont{L.~C.} \bibnamefont{Brunel}},
  \bibinfo{author}{\bibfnamefont{H.}~\bibnamefont{Kungl}}, \bibnamefont{and}
  \bibinfo{author}{\bibfnamefont{M.~J.} \bibnamefont{Hoffmann}},
  \bibinfo{journal}{Magn. Reson. Chem.} \textbf{\bibinfo{volume}{43}},
  \bibinfo{pages}{S166} (\bibinfo{year}{2005}{\natexlab{b}}).

\bibitem[{\citenamefont{Trasatti}(1990)}]{Tra90}
\bibinfo{author}{\bibfnamefont{S.}~\bibnamefont{Trasatti}},
  \bibinfo{journal}{Electrochimica Acta} \textbf{\bibinfo{volume}{35}},
  \bibinfo{pages}{269} (\bibinfo{year}{1990}).

\bibitem[{\citenamefont{Robertson and Chen}(1999)}]{RobChe99}
\bibinfo{author}{\bibfnamefont{J.}~\bibnamefont{Robertson}} \bibnamefont{and}
  \bibinfo{author}{\bibfnamefont{C.~W.} \bibnamefont{Chen}},
  \bibinfo{journal}{Appl. Phys. Lett.} \textbf{\bibinfo{volume}{74}},
  \bibinfo{pages}{1168} (\bibinfo{year}{1999}).

\bibitem[{\citenamefont{Lambeck and Jonker}(1986)}]{LamJon86}
\bibinfo{author}{\bibfnamefont{P.~V.} \bibnamefont{Lambeck}} \bibnamefont{and}
  \bibinfo{author}{\bibfnamefont{G.~H.} \bibnamefont{Jonker}},
  \bibinfo{journal}{J. Phys. Chem. Solids} \textbf{\bibinfo{volume}{47}},
  \bibinfo{pages}{453} (\bibinfo{year}{1986}).

\bibitem[{\citenamefont{Arlt}(1987)}]{Arl87}
\bibinfo{author}{\bibfnamefont{G.}~\bibnamefont{Arlt}},
  \bibinfo{journal}{Ferroelectrics} \textbf{\bibinfo{volume}{76}},
  \bibinfo{pages}{451} (\bibinfo{year}{1987}).

\bibitem[{\citenamefont{Carl and H\"ardtl}(1978)}]{CarHar78}
\bibinfo{author}{\bibfnamefont{K.}~\bibnamefont{Carl}} \bibnamefont{and}
  \bibinfo{author}{\bibfnamefont{K.}~\bibnamefont{H\"ardtl}},
  \bibinfo{journal}{Ferroelectrics} \textbf{\bibinfo{volume}{17}},
  \bibinfo{pages}{473} (\bibinfo{year}{1978}).

\bibitem[{\citenamefont{Takahashi}(1982)}]{Tak82}
\bibinfo{author}{\bibfnamefont{S.}~\bibnamefont{Takahashi}},
  \bibinfo{journal}{Ferroelectrics} \textbf{\bibinfo{volume}{41}},
  \bibinfo{pages}{143} (\bibinfo{year}{1982}).

\bibitem[{\citenamefont{Warren et~al.}(1996)\citenamefont{Warren, Pike,
  Vanheusden, Dimos, Tuttle, and Robertson}}]{WarPikVan96}
\bibinfo{author}{\bibfnamefont{W.~L.} \bibnamefont{Warren}},
  \bibinfo{author}{\bibfnamefont{G.~E.} \bibnamefont{Pike}},
  \bibinfo{author}{\bibfnamefont{K.}~\bibnamefont{Vanheusden}},
  \bibinfo{author}{\bibfnamefont{D.}~\bibnamefont{Dimos}},
  \bibinfo{author}{\bibfnamefont{B.~A.} \bibnamefont{Tuttle}},
  \bibnamefont{and}
  \bibinfo{author}{\bibfnamefont{J.}~\bibnamefont{Robertson}},
  \bibinfo{journal}{J. Appl. Phys.} \textbf{\bibinfo{volume}{79}},
  \bibinfo{pages}{9250} (\bibinfo{year}{1996}).

\bibitem[{\citenamefont{Ren}(2004)}]{Ren04}
\bibinfo{author}{\bibfnamefont{X.}~\bibnamefont{Ren}}, \bibinfo{journal}{Nature
  Mater.} \textbf{\bibinfo{volume}{3}}, \bibinfo{pages}{91}
  (\bibinfo{year}{2004}).

\bibitem[{\citenamefont{Robels et~al.}(1992)\citenamefont{Robels, Zadon, and
  Arlt}}]{RobZadArl92}
\bibinfo{author}{\bibfnamefont{U.}~\bibnamefont{Robels}},
  \bibinfo{author}{\bibfnamefont{C.}~\bibnamefont{Zadon}}, \bibnamefont{and}
  \bibinfo{author}{\bibfnamefont{G.}~\bibnamefont{Arlt}},
  \bibinfo{journal}{Ferroelectrics} \textbf{\bibinfo{volume}{133}},
  \bibinfo{pages}{163} (\bibinfo{year}{1992}).

\bibitem[{\citenamefont{Ghosez et~al.}(1997)\citenamefont{Ghosez, Gonze, and
  Michenaud}}]{GhoGonMic97}
\bibinfo{author}{\bibfnamefont{P.}~\bibnamefont{Ghosez}},
  \bibinfo{author}{\bibfnamefont{X.}~\bibnamefont{Gonze}}, \bibnamefont{and}
  \bibinfo{author}{\bibfnamefont{J.~P.} \bibnamefont{Michenaud}},
  \bibinfo{journal}{Ferroelectrics} \textbf{\bibinfo{volume}{194}},
  \bibinfo{pages}{39 } (\bibinfo{year}{1997}).

\bibitem[{\citenamefont{Cockayne and Burton}(2000)}]{CocBur00}
\bibinfo{author}{\bibfnamefont{E.}~\bibnamefont{Cockayne}} \bibnamefont{and}
  \bibinfo{author}{\bibfnamefont{B.~P.} \bibnamefont{Burton}},
  \bibinfo{journal}{Phys. Rev. B} \textbf{\bibinfo{volume}{62}},
  \bibinfo{pages}{3735} (\bibinfo{year}{2000}).

\bibitem[{\citenamefont{Cockayne}(2003)}]{Coc03}
\bibinfo{author}{\bibfnamefont{E.}~\bibnamefont{Cockayne}},
  \bibinfo{journal}{J. Eur. Ceram. Soc.} \textbf{\bibinfo{volume}{23}},
  \bibinfo{pages}{2375} (\bibinfo{year}{2003}).

\bibitem[{\citenamefont{Burns and Scott}(1973)}]{BurSco73}
\bibinfo{author}{\bibfnamefont{G.}~\bibnamefont{Burns}} \bibnamefont{and}
  \bibinfo{author}{\bibfnamefont{B.~A.} \bibnamefont{Scott}},
  \bibinfo{journal}{Phys. Rev. B} \textbf{\bibinfo{volume}{7}},
  \bibinfo{pages}{3088} (\bibinfo{year}{1973}).

\bibitem[{\citenamefont{Li et~al.}(1993)\citenamefont{Li, Grimsditch, Xu, and
  Chan}}]{LiGriXu93}
\bibinfo{author}{\bibfnamefont{Z.}~\bibnamefont{Li}},
  \bibinfo{author}{\bibfnamefont{M.}~\bibnamefont{Grimsditch}},
  \bibinfo{author}{\bibfnamefont{X.}~\bibnamefont{Xu}}, \bibnamefont{and}
  \bibinfo{author}{\bibfnamefont{S.-K.} \bibnamefont{Chan}},
  \bibinfo{journal}{Ferroelectrics} \textbf{\bibinfo{volume}{141}},
  \bibinfo{pages}{313} (\bibinfo{year}{1993}).

\bibitem[{\citenamefont{Kalinichev et~al.}(1997)\citenamefont{Kalinichev, Bass,
  Sun, and Payne}}]{KalBasSun97}
\bibinfo{author}{\bibfnamefont{A.~G.} \bibnamefont{Kalinichev}},
  \bibinfo{author}{\bibfnamefont{J.~D.} \bibnamefont{Bass}},
  \bibinfo{author}{\bibfnamefont{B.~N.} \bibnamefont{Sun}}, \bibnamefont{and}
  \bibinfo{author}{\bibfnamefont{D.~A.} \bibnamefont{Payne}},
  \bibinfo{journal}{J. Mater. Res.} \textbf{\bibinfo{volume}{12}},
  \bibinfo{pages}{2623} (\bibinfo{year}{1997}).

\bibitem []{Note1}
  \BibitemOpen
  \bibinfo {note} {While generalized gradient approximations (GGA) for the
    exhcange-correlation potential give often better agreement with
    experimental data than the local denisty approximation, this does
    not in general apply to ferroelectric perovskites. In the present
    case GGAs are observed to overestimate considerably the equilibrium
    volume, the axial ratio as well as the ferroelectric distortion. In
    particular for studies of defect properties the LDA is therefore
    often preferred (see e.g., Refs.~\onlinecite{ParCha98, MesEicKlo05,
      PoyCha00a, CocBur04, Par03, HeVan03, ZhaWuLu06,
      PoyCha99a, PoyCha00b}).}
  \BibitemShut {Stop}%

\bibitem []{Note2}
  \BibitemOpen
  \bibinfo {note} {Ideally, one would obtain the zero-temperature static
  dielectric tensor from a first-principles calculation. To the best of our
  knowledge, such data is not available for \text {PbTiO$_3$}, and its
  determination is a work in its own right (see e.g., Refs.~\protect
  \onlinecite{GhoGonMic97, CocBur00, Coc03}). We, therefore, resort here
  to experimental data. If one extrapolates the temperature dependent data from
  Ref.~\protect \onlinecite{BurSco73} to 0\protect \tmspace +\thinmuskip
  {.1667em}K, one obtains the zero-temperature, clamped dielectric constants
  $\epsilon _{11}^S=100$ and $\epsilon _{33}^S=34$ in reasonable agreement with
  (room-temperature) measurements (see Refs.~\protect \onlinecite
  {LiGriXu93} and \protect \onlinecite{KalBasSun97} for an overview of
  experimental measurements).}
  \BibitemShut {Stop}%

\bibitem []{Note3}
  \BibitemOpen
  \bibinfo {note} {Note, this correction does not take into account electronic
  relaxations which are obtained upon proper removal of self-interaction
  effects.}
  \BibitemShut {Stop}%


\end{thebibliography}
\end{document}